%% file: 0_Main.tex
\documentclass[acmtog]{acmart}
\acmSubmissionID{133}

\usepackage{booktabs} 
\usepackage{wrapfig}
\usepackage[ruled]{algorithm2e} 

\SetAlFnt{\small}
\SetAlCapFnt{\small}
\SetAlCapNameFnt{\small}
\SetAlCapHSkip{0pt}

\acmJournal{TOG}

\input{Commands.tex}

\begin{document}

\title{Differentiable Geodesic Distance for Intrinsic Minimization on Triangle Meshes}

\author{Yue Li}
\affiliation{%
  \institution{ETH Z{\"u}rich}
  \country{Switzerland}
}
\email{yue.li@inf.ethz.ch}

\author{Logan Numerow}
\affiliation{%
  \institution{ETH Z{\"u}rich}
  \country{Switzerland}
}
\email{lnumerow@inf.ethz.ch}

\author{Bernhard Thomaszewski}
\affiliation{%
  \institution{ETH Z{\"u}rich}
  \country{Switzerland}
}
\email{bthomasz@ethz.ch}

\author{Stelian Coros}
\affiliation{%
  \institution{ETH Z{\"u}rich}
  \country{Switzerland}
}
\email{stelian.coros@inf.ethz.ch}


\begin{abstract}
Computing intrinsic distances on discrete surfaces is at the heart of many minimization problems in geometry processing and beyond. Solving these problems is extremely challenging as it demands the computation of on-surface distances along with their derivatives. We present a novel approach for intrinsic minimization of distance-based objectives defined on triangle meshes. Using a variational formulation of shortest-path geodesics, we compute first and second-order distance derivatives based on the implicit function theorem, thus opening the door to efficient Newton-type minimization solvers. We demonstrate our differentiable geodesic distance framework on a wide range of examples, including geodesic networks and membranes on surfaces of arbitrary genus, two-way coupling between hosting surface and embedded system, differentiable geodesic Voronoi diagrams, and efficient computation of Karcher means on complex shapes. Our analysis shows that second-order descent methods based on our differentiable geodesics outperform existing first-order and quasi-Newton methods by large margins.
\end{abstract}

%

\setcopyright{acmlicensed}
\acmJournal{TOG}
\acmYear{2024} \acmVolume{43} \acmNumber{4} \acmArticle{91} \acmMonth{7}\acmDOI{10.1145/3658122}


\maketitle
\input{1_Introduction}
\input{2_RelatedWork}
\input{3_Method}
\input{4_Results}

\input{5_Conclusion}
\begin{acks}
We thank Peiyuan Xie for his assistance with the visualization, and the anonymous reviewers for their valuable feedback. This work was supported by the European Research Council (ERC) under the European Union’s Horizon 2020 research and innovation program (grant agreement No. 866480), and the Swiss National Science Foundation through SNF project grant 200021\_200644.
\end{acks}
\bibliographystyle{ACM-Reference-Format}
\bibliography{reference}
\input{Appendix.tex}

\end{document}

%% file: Commands.tex
\usepackage{xspace}
\makeatletter
\DeclareRobustCommand\onedot{\futurelet\@let@token\@onedot}
\def\@onedot{\ifx\@let@token.\else.\null\fi\xspace}
 
\def\ie{\emph{i.e}\onedot}

\def\wrt{w.r.t\onedot} 
\def\etal{\emph{et al}\onedot}
\makeatother

\usepackage[makeroom]{cancel}

\newcommand{\bA}{\mathbf{A}}
\newcommand{\bB}{\mathbf{B}}
\newcommand{\bC}{\mathbf{C}}
\newcommand{\bD}{\mathbf{D}}

\newcommand{\bF}{\mathbf{F}}

\newcommand{\bH}{\mathbf{H}}

\newcommand{\bc}{\mathbf{c}}

\newcommand{\be}{\mathbf{e}}
\newcommand{\bff}{\mathbf{f}}
\newcommand{\bg}{\mathbf{g}}

\newcommand{\bp}{\mathbf{p}}

\newcommand{\bs}{\mathbf{s}}
\newcommand{\bt}{\mathbf{t}}

\newcommand{\bv}{\mathbf{v}}
\newcommand{\bx}{\mathbf{x}}

\newcommand{\bq}{\mathbf{q}}
\newcommand{\bw}{\mathbf{w}}
\newcommand{\by}{\mathbf{y}}
\newcommand{\bZero}{\mathbf{0}}

\newcommand{\tran}{\mathsf{T}}
\newcommand{\dd}{\mathrm{d}}

\newcommand{\tbx}{\tilde{\mathbf{x}}}

\DeclareMathOperator*{\argmin}{argmin}

\newcommand*{\argminl}{\argmin\limits}

\newif\ifcomment
\commenttrue

%% file: 1_Introduction.tex
\section{Introduction}
Computing intrinsic distances on triangle meshes is a fundamental task in geometry processing. 
From biological films on surfaces, to the fascia enveloping our muscles, and to tight-fitting clothing---there are countless examples of variational problems where the task is to minimize lengths on discrete manifolds.
Solving such problems on triangle meshes requires the computation of geodesic distances on surfaces along with their derivatives.
\begin{figure}[ht]
    \centering
    \includegraphics[width=\linewidth]{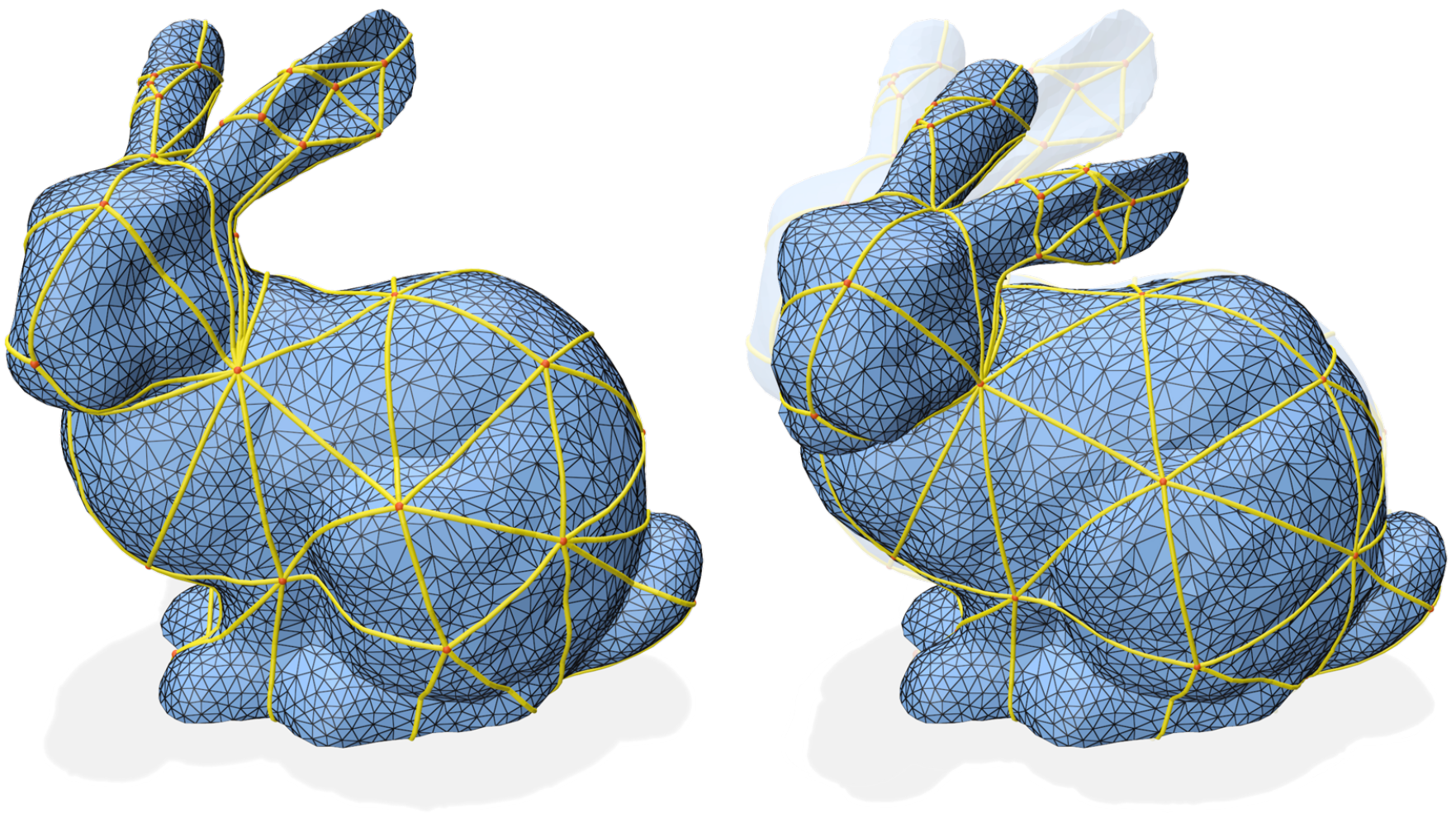}
    \caption{Embedded two-way coupling. We simulate a geodesic elastic network embedded in the surface of a deformable bunny, modeled with solid finite elements. Our analytical geodesic distance derivatives allow us to use Newton's method for simulating the deformations induced by tightening the network.}
    \label{fig:teaser}
\end{figure}
While gradients are readily computed, the convergence of first-order methods is generally poor, in particularly for embedded elasticity problems with stiff connections. 
Quasi-Newton methods such as L-BFGS may offer some acceleration, but as we show in our analysis, they still converge slowly, especially for stiff problems.
\par
In this work, we present a novel approach for intrinsic minimization of arbitrary objectives based on geodesic lengths on piece-wise linear surfaces. 
Using a variational form of shortest-path geodesics, we obtain distance derivatives in closed form using the implicit function theorem. We show that first and second derivatives can be simplified to enable efficient computation, allowing us to leverage powerful Newton-type methods for simulation.
Our approach is underpinned by the key observation that, although geodesic paths on triangle meshes are generally not smooth functions of their endpoints, geodesic distance is at least $C^0$-continuous everywhere---and infinitely smooth if geodesics are unique. Our analysis reveals two types of gradient discontinuities, one of which can be resolved through mollification while the other one does not occur close to minima and can therefore be ignored.

We demonstrate our differentiable geodesic distance framework on a large and diverse set of minimization problems. Specifically, we construct embedded elastic networks and membranes based on geodesic springs and triangle finite elements, respectively.
We present simulation examples on shapes with varying topologies, including two-way coupling with the hosting surface. 
We furthermore show how our formulation enables differentiable on-surface Voronoi diagrams, whose site locations can be optimized for various objectives. 
Finally, we demonstrate that our simulation framework can be used to compute Karcher means on arbitrary surfaces in efficient and highly accurate ways. 

%% file: 2_RelatedWork.tex
\section{Related Work}
\paragraph{Discrete Geodesic Distance}
Geodesic distance has been extensively studied in the geometry processing community~\cite{Crane:2020:SAG,peyre2010geodesic,bose2011survey}. Many methods have been developed for calculating exact geodesic distances~\cite{mitchell1987discrete,chen1990shortest,qin2016fast,sharp2020you,liu2017optimization} as well as fast approximations~\cite{crane2013geodesics,belyaev2015variational,surazhsky2005fast,10.1145/3588432.3591523,belyaev2020admm,kimmel1998computing,ying2013saddle,trettner2021geodesic,pang2023learning}. While approximations through the solution of smooth energy functions offer advantages in terms of differentiability, accurately tracing the geodesic path from the distance field is a challenging task in itself~\cite{Crane:2020:SAG}. We use an exact geodesic distance computation from which geodesic paths can be constructed. Although the numerical properties of discrete geodesics have been studied intensively, we are not aware of any second-order method for minimizing geodesics-based objectives. 
To construct such a method, we propose a differentiable geodesic distance formulation that is almost everywhere $C^2$-continuous and exhibits robust convergence when integrated into Newton-type minimization algorithms.
\paragraph{Intrinsic Geometry Processing}
The computer graphics community has made great strides in intrinsic geometry processing~\cite{liu2017constructing,bobenko2007discrete,xin2011isotropic,fisher2006algorithm,Sharp:2019:NIT,sharp2020laplacian,gillespie2021integer,liu2023surface}. Intrinsic triangulations facilitate tasks such as Delaunay refinement~\cite{Sharp:2019:NIT}, mesh simplification~\cite{liu2023surface}, and construction of differential operators on non-manifold meshes~\cite{sharp2020laplacian}. 
Similar to intrinsic triangulations~\cite{Sharp:2019:NIT}, our method uses endpoints as only degrees of freedom---connecting geodesic edges are defined implicitly and reconstructed on demand. Unlike existing work, our approach enables the computation of first and second derivatives of geodesic distances in closed form, thus opening the door to efficient second-order minimization algorithms.
%
\paragraph{Structural Curve Networks}
Structural curve networks have gained increasing attention in the fields of computer graphics and robotics due to their aesthetic appeal and practical utility.
For instance, Schumacher~\etal~\shortcite{schumacher2018mechanical} characterize the mechanical behavior of different families of tiling patterns,
and Li~\etal~\shortcite{li2022programmable} explore the direction-dependent stiffness of 3D-printed weave structures.
Another line of research focuses on the structural stability of curve networks~\cite{zehnder2016designing,perez2015design,miguel2016computational,ren20213d,neveu2022stability,liu2021computational}.
Our work is similar in the sense that we operate on curves embedded within surfaces. Instead of representing these curves explicitly, however, they are represented implicitly as shortest geodesic paths. Our differentiable geodesic distance formulation allows us to solve intrinsic minimization problems defined on these implicit curve networks.
%
%
\paragraph{Embedded Simulation}
Many physical phenomena can be describ\-ed through partial differential equations on Riemannian manifolds. Examples include swirl dynamics on soap bubbles~\cite{yang2019real,huang2020chemomechanical}, skin sliding~\cite{li2013thin}, tight-fitting clothing \cite{montes2020computational}, and elastic curve networks embedded in curved surfaces \cite{zehnder2016designing}.
One approach to this embedded elasticity problem is to use 3D spline curves whose control vertices are constrained to lie on the surface~\cite{lee2001geometric,hofer2004energy,wallner2005fair}. 
Eulerian-on-Lagrangian methods are an alternative representation for simulating constrained motion of deformable bodies~\cite{fan2013eulerian}, rods~\cite{sueda2011large,li2022programmable}, and cloth~\cite{cirio2014yarn,cirio2015efficient,weidner2018eulerian}.
Within this context, Li~\etal~\shortcite{li2013thin} simulate skin sliding using texture-like material coordinates as Eulerian degrees of freedom. In order for this approach to work, however, a texture atlas of the underlying surface with quasi-isometric charts is required. 
Montes~\etal~\shortcite{montes2020computational} propose a Lagrangian-on-Lagrangian approach that combines subdivision surfaces with embedded triangle meshes to allow for smooth sliding of skin-tight clothing across the underlying body. However, they approximate on-surface lengths using Euclidean distances. Consequently, surface and cloth discretizations must have adequate resolutions to limit approximation error. 
Our simulation examples with embedded elastic networks likewise use a Lagrangian-on-Lagrangian representation. 
However, we avoid resolution dependence by computing exact geodesic distances on triangle meshes.
%

%% file: 3_Method.tex
\section{Background and Challenges}
Our goal is to minimize distance-based functions on triangle meshes with second-order optimizers. Naturally, this requires the distance metric to be sufficiently smooth and differentiable. While Euclidean distance satisfies these criteria, depending on the resolution of the embedded mesh, it can deviate from the actual distance between two surface points, \ie, the geodesic distance, to different degrees. This deviation can lead to undesired local minima, hindering the optimization process (see Sec.~\ref{sec:ablation_study}). We therefore opt to use geodesic distances for simulation.
Before describing our method in detail, we first briefly discuss the smoothness of geodesic distance. 

\subsection{Geodesic Distance}
In the continuous setting, geodesics are locally shortest paths between two points on a surface. These paths are also straightest, \ie, they exhibit zero geodesic curvature. However, in the discrete setting with non-smooth surfaces, these two definitions---straightest and shortest path---are not equivalent and will generally lead to different geodesic paths. 
The shortest path definition is a natural choice for boundary value problems with known endpoints, while the straightest-path definition is best suited for initial value problems with a given starting point and tangent direction. We refer to Crane \etal~\shortcite{Crane:2020:SAG} for an in-depth discussion of this subject. In this work, we use shortest-path geodesics, which can be computed by minimizing a convex quadratic potential.
To use this distance metric in a minimization algorithm, we must first understand its smoothness properties.

\subsection{Smoothness of Geodesic Distance}
The geodesic distance between two points is a continuous function of the points everywhere on the surface. The geodesic path, however, is not always unique and can change abruptly for smooth motion of the endpoints. The distance derivative is necessarily discontinuous in these singular configurations. It is worth mentioning that this discontinuity exists even for smooth surfaces and does not originate from discretization.

\paragraph{Continuous Setting.} 
\label{sec:geo_con}
To understand this discontinuity, consider a surface patch containing a bump and a geodesic whose endpoints lie on opposite sides of the bump (see inset figure). As we translate the upper endpoint to the right, the geodesic path reaches a singular point where it suddenly \textit{flips} to the right.
At the singularity, there exist two paths on different sides of the bump that have the same geodesic distance. A local perturbation of the endpoint can easily lead to flipping of the shortest geodesic. The collection of points 
\setlength{\columnsep}{3pt}
\begin{wrapfigure}{r}{0.25\textwidth} 
    \vspace{-10pt}
    \centering
    \includegraphics[width=\linewidth]{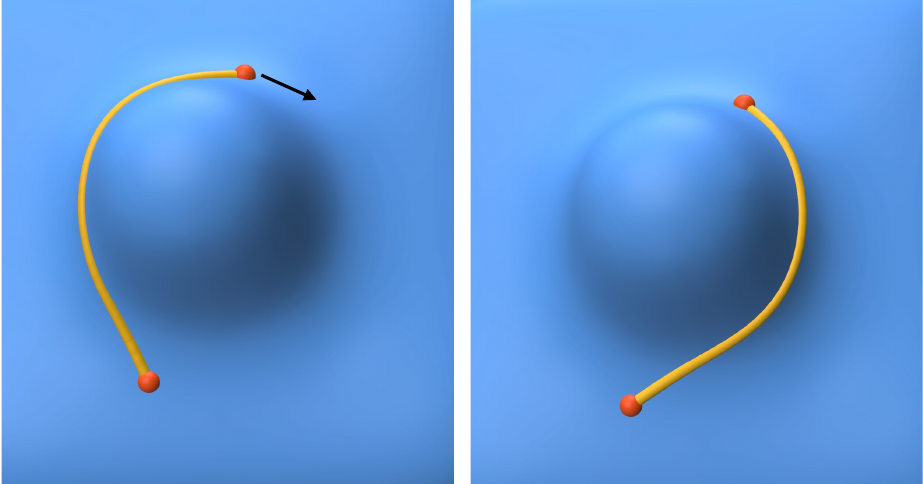}
    \label{fig:con_cur}
    \vspace{-20pt}
\end{wrapfigure}
for which geodesics are not unique is referred to as the \textit{cut locus}. The shortest geodesic path from point $a$ to point $b$ is discontinuous as $b$ passes through the cut locus of $a$, and along the cut locus, the gradient of the path is undefined. Although this discontinuity in the distance gradient exists even in the continuous setting, it does not prevent gradient-based minimization since points on the cut locus locally \textit{maximize} the geodesic distance from the source point.

\paragraph{Discrete Setting.}
Geodesic distance remains continuous in the discrete setting, but geodesic paths behave somewhat differently compared to the smooth setting. Special care is required when geodesics pass close to mesh vertices, which we classify into three categories depending on their discrete Gaussian curvature: spherical vertices (positive curvature), hyperbolic vertices (negative curvature), and planar vertices (zero curvature). 
\begin{figure}[h]
    \centering
    \includegraphics[width=\linewidth]{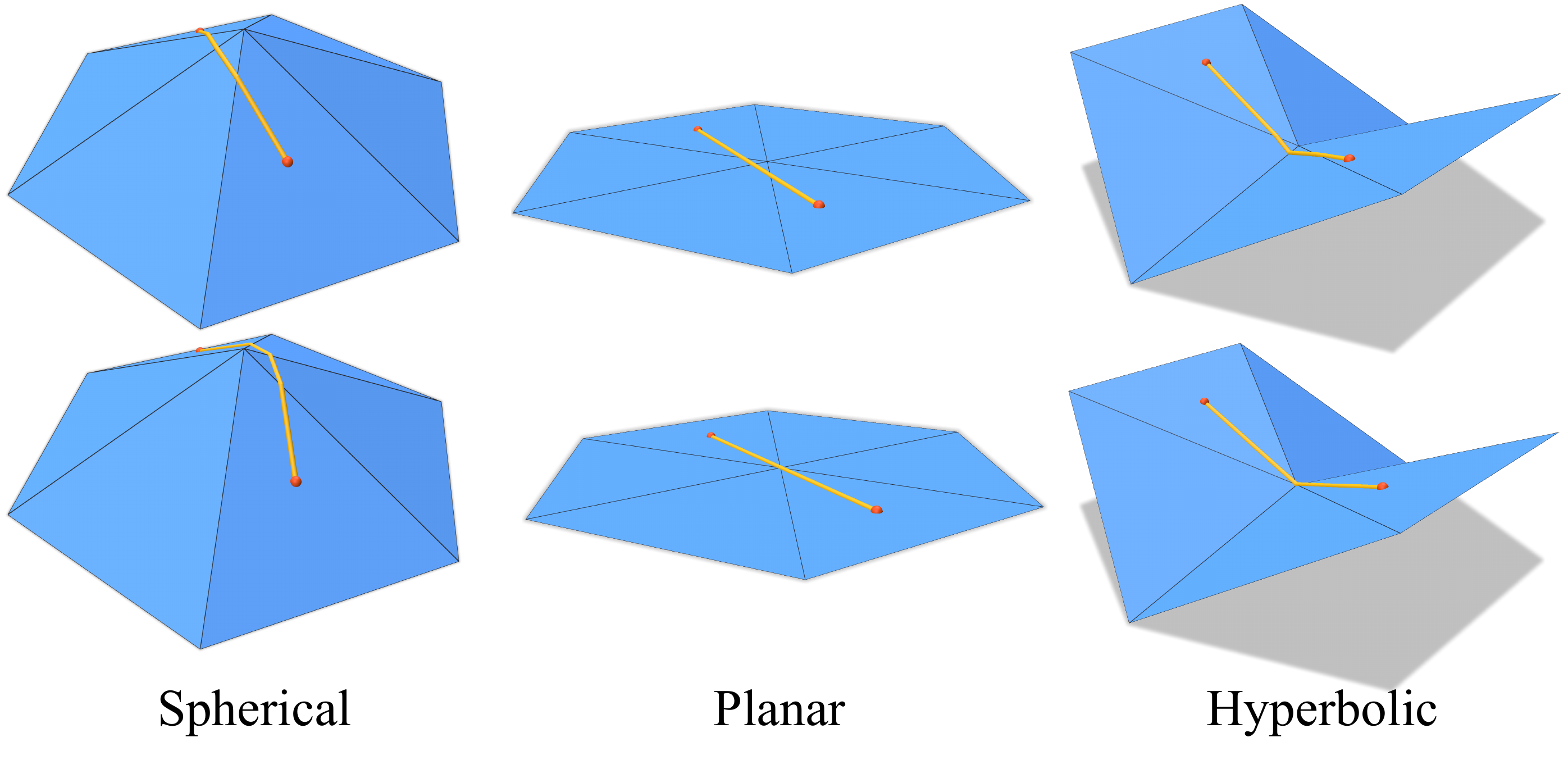}
    \caption{Behavior of geodesic paths passing over different types of vertices. 
    Fixing one endpoint of a geodesic, we translate the other such that the path moves across the center vertex. Whereas the geodesic passes through the center vertex in the planar and hyperbolic case, it jumps over the spherical vertex to avoid the local distance maximum.}
    \label{fig:dis_cur}
\end{figure}
In the vicinity of planar vertices, the geodesic is a straight line and its length reduces to the Euclidean distance.
For spherical vertices, however, a geodesic cannot pass through the vertex as there always exists a shorter path going around it (see Fig.~\ref{fig:dis_cur}). 
Since geodesic paths vary discontinuously around spherical vertices, geodesic distance is only $C^0$-continuous. Fortunately, paths through spherical vertices are length-maximizing and are thus avoided during distance minimization. 
Lastly, geodesic paths can continuously pass through hyperbolic vertices, but the change in local tangent direction leads to only $C^1$-continuity in these cases. In Sec.~\ref{sec:dif_geo} we show that $C^2$-continuity can be restored by adding a suitable mollifier. 
\subsection{Challenges}
Our approach builds on the observation that, while geodesic paths are not generally continuous functions of their endpoints, their lengths vary continuously and can thus be used for gradient-based minimization. 
Computing the required derivatives, however, is no trivial task. Geodesics on triangle meshes generally span multiple faces and intersect with the edges of the hosting surface. Explicitly tracking a varying number of intersection variables is cumbersome. Using endpoints as degrees of freedom circumvents this issue, but intersection points are then functions of the endpoints, adding another layer of complexity to the computation of first and second derivatives. In fact, we are not aware of any existing work that computes the analytical derivatives of geodesic distance on triangle meshes.
In the following, we develop a differentiable geodesic distance formulation whose first and second derivatives can be computed efficiently. Since our formulation requires only intrinsic quantities, we refer to it as \textit{intrinsic minimization}.
\section{Intrinsic Minimization}
We develop a formulation for differentiable geodesic distance on triangle meshes that uses geodesic endpoints as the only explicit variables. The intersection points on the geodesic path are implicitly defined through equilibrium conditions of a shortest-path energy functional (Sec.~\ref{sec:dif_geo}). We show that an intrinsic simulation framework driven by geodesic distance can be formulated on this basis (Sec.~\ref{sec:int_rep}) and the required derivatives can be obtained in simple forms using sensitivity analysis and geometric insights. We extend our distance-driven formulation from geodesic edges to triangles such as to construct directionally-continuous deformation energies (Sec.~\ref{sec:elastic_triangles}). We furthermore elaborate on two-way coupling effects between geodesic networks and their embedding surfaces (Sec.~\ref{sec:two_way}). Finally, beyond simulating embedded elasticity, we generalize the intrinsic minimization paradigm to differentiable geodesic Voronoi diagrams (Sec.~\ref{sec:diff_gvd}).
%
\subsection{Differentiable Geodesic Distance}
\label{sec:dif_geo}
\setlength{\columnsep}{2pt}
\begin{wrapfigure}{r}{0.17\textwidth} 
    \centering
    \vspace{-20pt}
    \includegraphics[width=\linewidth]{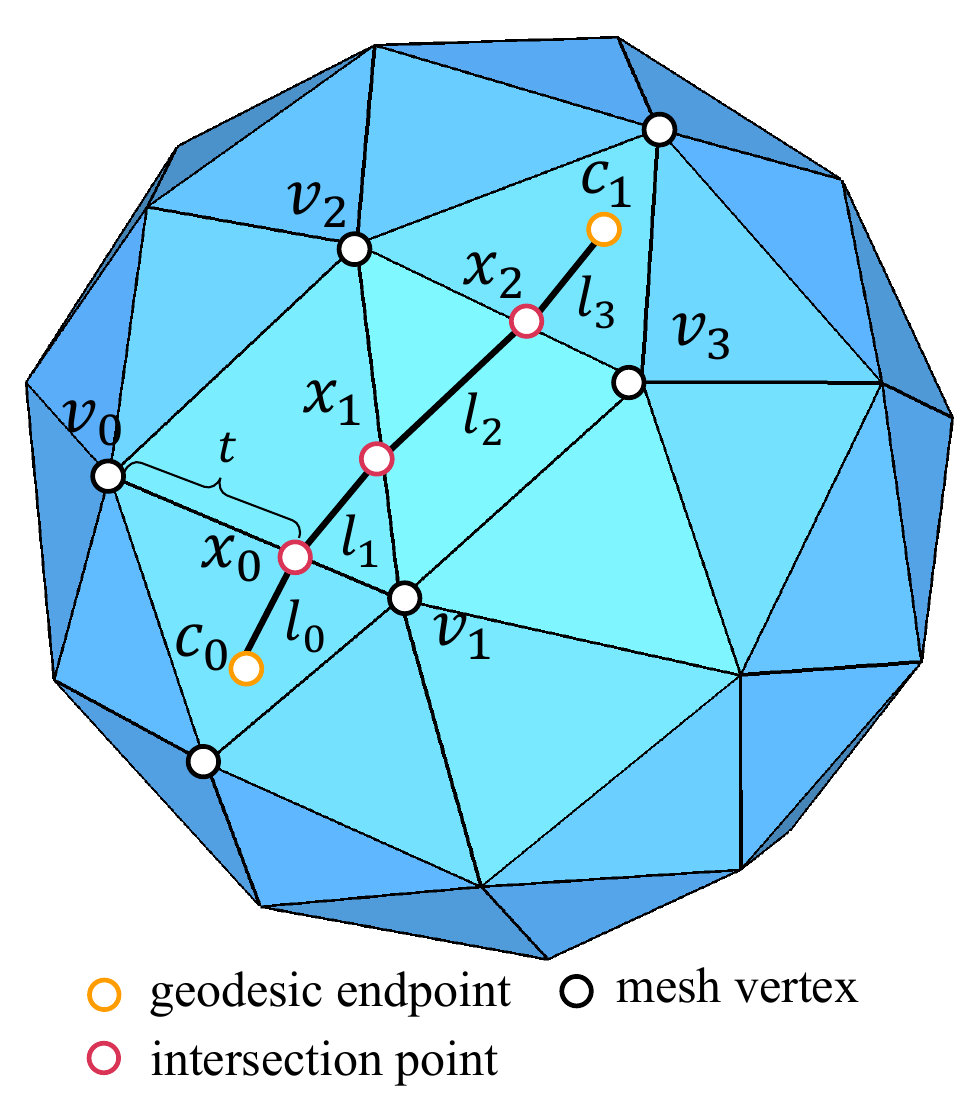}
    \vspace{-20pt}
    \label{fig:geodesic_notation}
\end{wrapfigure}
Our approach builds on the robust MMP algorithm~\cite{mitchell1987discrete} to compute exact geodesic paths on triangle meshes. The endpoints of geodesic paths are represented using barycentric coordinates $\bw$ of the hosting triangle mesh. These endpoints are the only simulation degrees of freedom of a given geodesic path. The spatial coordinates of the endpoints $\bc$, and the intersections $\bx$ between the path and edges of the hosting mesh, are computed when either endpoint is moved (see inset). We use a scalar parameter $t_i$ to denote the position of an intersection point $\bx_i$ along a corresponding mesh edge $e_{jk}$, 
\begin{equation}\label{eqn:intersection_pt_parameterization}
    \bx_i = \bv_j + t_i(\bv_k - \bv_j) \ ,
\end{equation}
where $\bv$ are mesh vertices.
We compute the geodesic distance $g$ between two points, $\bc_0$ and $\bc_1$, on the surface mesh as the sum of the lengths of the line segments $l_i$ comprising the geodesic path,
\begin{equation}
    g(\bc_0, \bc_1) = \sum_i l_i(\bc_0, \bc_1, \bx(\bc_0, \bc_1)) \ .
\end{equation}
The relationship of the simulation variables $\bw$ and geodesic endpoints $\bc$ are given explicitly by interpolating barycentric coordinates. The relationship of the intersection variables $\bt$ are given implicitly by the algorithm for computing exact geodesics.
All involved quantities ultimately depend only on the simulation variables $\bw$ and we therefore write
\begin{equation}
\label{eqn:geodesic}
    g(\bc,\bx)= g(\bc(\bw), \bx(\bt(\bc(\bw))) \ .
\end{equation}
To enable Newton-type solvers for intrinsic minimization, we must compute the first and second derivatives of the above expression. The first derivative is given by
\begin{equation}
    \frac{\dd g}{\dd \bw} = \frac{\partial g}{\partial \bc}^{\tran} \frac{\partial \bc}{\partial \bw} + \frac{\partial g}{\partial \bx}^{\tran} \frac{\partial \bx}{\partial \bt} \frac{\partial \bt}{\partial \bc} \frac{\partial \bc}{\partial \bw} \ ,
\end{equation}
and the second derivative follows as
\begin{equation}
\label{eq:HessianFull}
\begin{aligned}
    & \frac{\dd^2 g}{\dd \bw^2} = \frac{\partial \bc}{\partial \bw}^{\tran} \frac{\partial^2 g}{\partial \bc^2} \frac{\partial \bc}{\partial \bw} + \frac{\partial \bc}{\partial \bw}^{\tran} \frac{\partial^2 g}{\partial \bc \partial \bx} \frac{\partial \bx}{\partial \bt} \frac{\partial \bt}{\partial \bc}\frac{\partial \bc}{\partial \bw} + 
    \sum_i \frac{\partial g}{\partial c_i} \frac{\partial^2 c_i}{\partial \bw^2} \\
    & + \left(\frac{\partial \bx}{\partial \bt} \frac{\partial \bt}{\partial \bc}\frac{\partial \bc}{\partial \bw} \right)^{\tran} \left( \frac{\partial^2 g}{\partial \bc \partial \bx}\frac{\partial \bc}{\partial \bw} + \frac{\partial^2 g}{\partial \bx^2} \frac{\partial \bx}{\partial \bt} \frac{\partial \bt}{\partial \bc}\frac{\partial \bc}{\partial \bw} \right) \\
    & + 
    \left(\frac{\partial \bt}{\partial \bc} \frac{\partial \bc}{\partial \bw} \right)^{\tran}\sum_i \frac{\partial g}{\partial x_i} \frac{\partial^2 x_i}{\partial \bt^2} \left(\frac{\partial \bt}{\partial \bc} \frac{\partial \bc}{\partial \bw} \right)\\
    &+ 
    \frac{\partial \bc}{\partial \bw}^{\tran} \left( \sum_i \frac{\partial g}{\partial \bx} \frac{\partial \bx}{\partial t_i} \frac{\partial^2 t_i}{\partial \bc^2} \right)\frac{\partial \bc}{\partial \bw} + 
    \sum_i \frac{\partial g}{\partial \bx} \frac{\partial \bx}{\partial \bt}\frac{\partial \bt}{\partial c_i} \frac{\partial^2 c_i}{\partial \bw^2}\ .
\end{aligned}
\end{equation}
While computing all of these terms is feasible, we show that both expressions can be significantly simplified using geometric insight.
\par
The shortest geodesic is a local minimizer of distance, giving rise to the first-order optimality condition
\begin{equation}
\label{eqn:implicit_geodesic_zero}
    \frac{\dd g}{\dd \bt} = \frac{\partial g}{\partial \bx}^{\tran} \frac{\partial \bx}{\partial \bt} = \bZero \ .
\end{equation}
If the interior vertices $\bx$ satisfy this optimality condition, perturbations $\delta \bt$ to the intersection variables will not change length to first order: moving \textit{along} the path changes adjacent segment lengths, but these changes sum to zero; moving \textit{orthogonal} to the path simply rotates the segments, which does not change their lengths to first order. 
We can therefore simplify the total derivative to
\begin{equation}
    \frac{\dd g}{\dd \bw} = \frac{\partial g}{\partial \bc}^{\tran} \frac{\partial \bc}{\partial \bw}\ .
\end{equation}
This expression is readily evaluated since it only requires the gradient of length for the first and last segments and the (constant) Jacobian $\frac{\partial \bc}{\partial \bw}$ of the barycentric interpolation function. 
\par
\begin{figure}[h]
    \centering
    \includegraphics[width=\linewidth]{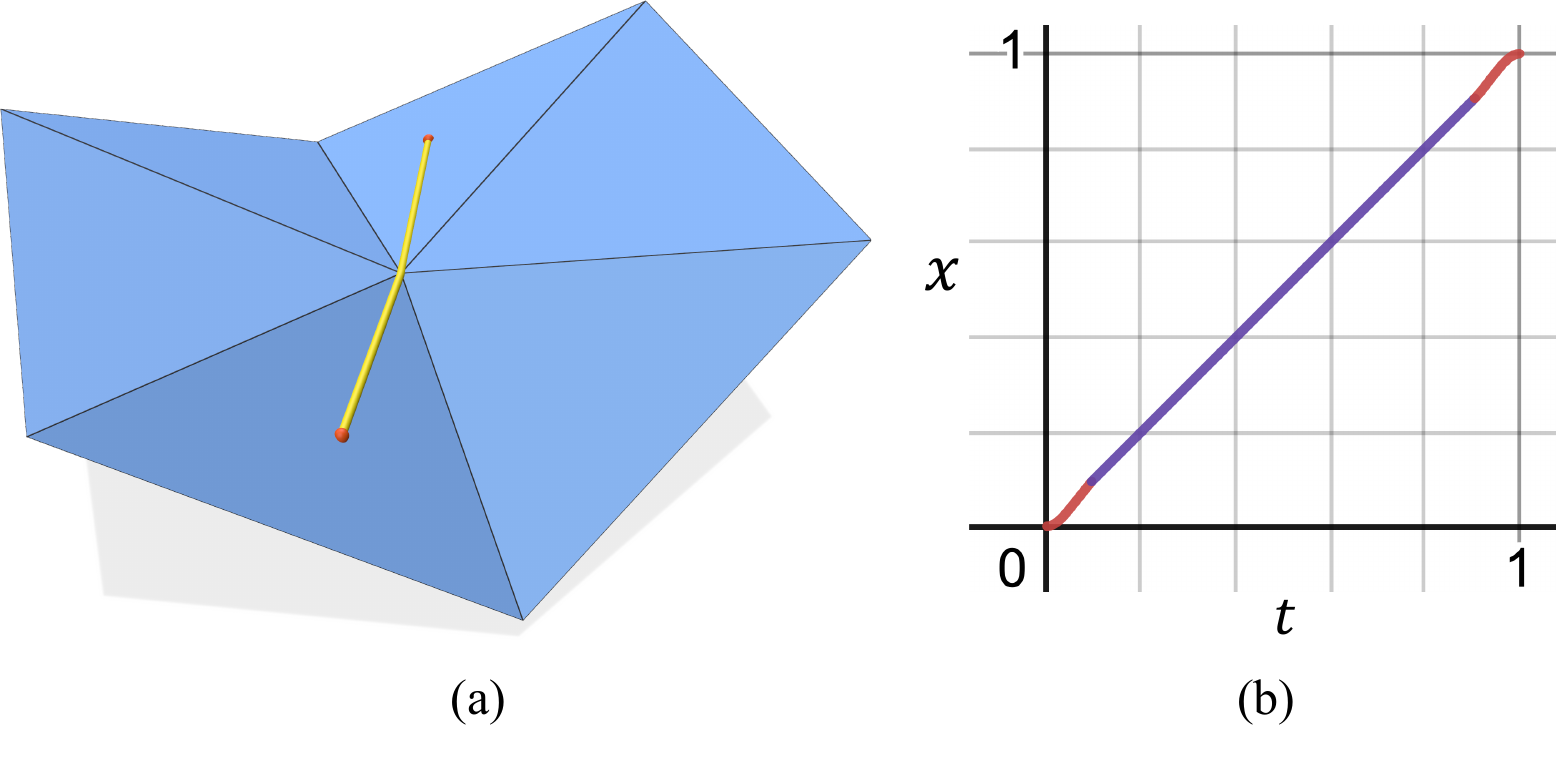}
    \caption{Mollification for edge intersection points. We smoothly blend linear edge intersection trajectories $\bx(t)$ with cubic functions (b) to achieve $C^2$-continuity when geodesics pass through mesh vertices (a). }
    \label{fig:hyperbolicDiscontinuity}
\end{figure}

Similarly, the Hessian can also be greatly simplified using implicit differentiation. Note that all terms can be evaluated algebraically except for $\frac{\partial \bt}{\partial \bc}$, \ie, the derivatives of the edge intersection points with respect to the endpoints of the geodesic. 
Computing this derivative explicitly requires differentiation through the entire pipeline of the algorithm used for computing the exact geodesics. Apart from the complexity of differentiating such a code, this stra\-te\-gy must fail for cases in which geodesics are not unique and, hence, path derivatives do not exist. Our key insight is that this problem can be entirely avoided by implicit differentiation. 
We begin by differentiating both sides of (\ref{eqn:implicit_geodesic_zero}) \wrt $\bc$, 
\begin{equation}
\label{eqn:dgdtdc}
    \frac{\partial \bx}{\partial \bt}^{\tran} \left( \frac{\partial^2 g}{\partial \bc \partial \bx} + \frac{\partial^2 g}{\partial \bx^2} \frac{\partial \bx}{\partial \bt} \frac{\partial \bt}{\partial \bc} \right) = \bZero\ .
\end{equation}
To obtain $\frac{\partial \bt}{\partial \bc}$, we rearrange and solve the small linear system
\begin{equation}
\label{eqn:solve_dtdc}
    \left(\frac{\partial \bx}{\partial \bt}^{\tran}\frac{\partial^2 g}{\partial \bx^2}\frac{\partial \bx}{\partial \bt}\right) \frac{\partial \bt}{\partial \bc} = -\frac{\partial \bx}{\partial \bt}^{\tran}\frac{\partial^2 g}{\partial \bc \partial \bx}\ .
\end{equation}
The effective size of this $m \times m$ system depends on the number $m$ of intersections between a given geodesic and the hosting mesh.
Multiplying (\ref{eqn:dgdtdc}) by $(\frac{\partial \bt}{\partial \bc}\frac{\partial \bc}{\partial \bw})^{\tran}$ from the left and by $\frac{\partial \bc}{\partial \bw}$ from the right gives
\begin{equation}
\label{eqn:simplify_dgdtdc}
\begin{aligned}
    \left( \frac{\partial \bx}{\partial \bt}\frac{\partial \bt}{\partial \bc}\frac{\partial \bc}{\partial \bw} \right)^{\tran} \left( \frac{\partial^2 g}{\partial \bc \partial \bx}\frac{\partial \bc}{\partial \bw} + \frac{\partial^2 g}{\partial \bx^2} \frac{\partial \bx}{\partial \bt}\frac{\partial \bt}{\partial \bc}\frac{\partial \bc}{\partial \bw} \right) &= \bZero\ .
\end{aligned}
\end{equation}
Substituting (\ref{eqn:implicit_geodesic_zero}) and (\ref{eqn:simplify_dgdtdc}) into (\ref{eq:HessianFull}) and removing second derivatives of linear terms, we obtain the substantially simplified expression
\begin{equation}
\label{eqn:d2gdw2_final}
    \frac{\partial^2 g}{\partial \bw^2} = \frac{\partial \bc}{\partial \bw}^{\tran} \frac{\partial^2 g}{\partial \bc^2} \frac{\partial \bc}{\partial \bw} + \frac{\partial \bc}{\partial \bw}^{\tran} \frac{\partial^2 g}{\partial \bc \partial \bx} \frac{\partial \bx}{\partial \bt} \frac{\partial \bt}{\partial \bc}\frac{\partial \bc}{\partial \bw}\ .
\end{equation}
\paragraph{Discontinuities \& Mollification}
With the expression for second derivatives in hand, we must still address potential discontinuities when intersection points $\bx(\bt)$ approach mesh vertices. 
Depending on to which side of a mesh vertex the geodesic passes, it will intersect different mesh edges. The function $\bx(t)$ is therefore ill-defined when the geodesic path coincides with a mesh vertex, \ie, when $t=0$ or $t=1$ for individual edges, and the optimality condition (Eq.~\ref{eqn:implicit_geodesic_zero}) does not hold.
This discontinuity hinders convergence, as shortest geodesics tend to pass through hyperbolic vertices (Fig. \ref{fig:hyperbolicDiscontinuity}\textit{a}), and the total distance energy is not $C^2$-continuous. 
We resolve this issue by adding a mollifier to $\bx(t)$ such that $\frac{d\bx(t)}{dt}$ smoothly vanishes at vertex intersections.
\par
A similar discontinuity appears when endpoints of geodesics coincide with vertices of the hosting mesh. We therefore apply the same mollifier to the barycentric coordinates of the endpoints. 
%
\paragraph{Smooth Mollifier}
We smoothly blend the linear intersection point parameterization (\ref{eqn:intersection_pt_parameterization}) with two cubic functions (see Fig.~\ref{fig:hyperbolicDiscontinuity}\textit{b}) such that the derivatives vanish when the geodesic curve passes through mesh vertices, \ie, when $t=0$ or $t=1$. The mollifier function is defined as 
\begin{equation}
    \label{eqn:mollifier}
    \hat{t}(t)=
\begin{cases}
    -\frac{t^3}{\epsilon^2} + \frac{2t^2}{\epsilon} &0 \leq t< \epsilon, \\
    t & \epsilon \leq t \leq 1 - \epsilon \\
    -\frac{(t-1)^3}{\epsilon^2} + \frac{2(t-1)^2}{\epsilon} + 1 & 1 - \epsilon < t \leq 1  \ ,
\end{cases}
\end{equation}
where $\epsilon=10^{-6}$ defines the smoothing length. This mollifier is  $C^1$-continuous along an edge, and as shown in Sec.~\ref{sec:ablation_study}, significantly accelerates the convergence of our Newton solver.
\subsection{Elastic Geodesic Networks}
\label{sec:int_rep}
With our differentiable geodesic distance formulation established, we now turn to energy minimization.
As a first example, we consider elastic curve networks comprised of geodesic springs. The total energy of this system is
\begin{equation}
    E_\mathrm{network}(\bw) = \sum_i (g_i(\bw) - \bar{g}_i)^2 \ ,
\end{equation}
where $i$ runs over all geodesic springs, whose current and rest lengths we denote as $g_i$ and $\bar{g}_i$, respectively.
We minimize this energy using Newton's method with a standard backtracking line search. The search direction per iteration is computed by solving the linear system
\begin{equation}
    \bH(\bw) \Delta \bw = -\by(\bw)\ ,
\end{equation}
where $\by(\bw)$ and $\bH(\bw)$ are the gradient and Hessian of the objective function. We use adaptive diagonal regularization~\cite{nocedal1999numerical} to ensure that the Hessian is positive definite.
Since all simulation variables are barycentric coordinates, the resulting search direction yields endpoint updates expressed in the barycentric coordinates of the corresponding mesh triangles. Inevitably, endpoints will move across edges and vertices into neighboring triangles, which requires updating the search direction to the new coordinate system. 
To this end, we use the approach by Sharp~\etal~\shortcite{Sharp:2019:NIT} for tracing  straightest geodesics~\cite{polthier2006straightest} and update local coordinates as
\begin{equation}
\begin{aligned}
\label{eqn:tracing}
    \bw^{j+1} &= \bw^j + \alpha \cdot \text{tracing}(\bw^j, \Delta \bw) \ ,
\end{aligned}
\end{equation}
where $\alpha$ is a step size determined by line search to ensure monotonic decrease in energy. 
\paragraph{Karcher Means}
Using a slightly different formulation, our method can be readily extended to compute so-called \textit{Karcher means} on triangle meshes. Consider a special case of a geodesic spring network, where a point $\bp$ is connected to several anchored points $\bx$ on a surface. The energy in this case amounts to the variational formulation for the Karcher mean,
\begin{equation}
\label{eqn:karcher}
    E_\mathrm{Karcher}(\bw) = \frac{1}{2N} \sum_{i=1}^{N} g(\bp(\bw),\bx_i)^2 \ .
\end{equation}
Sharp~\etal~\shortcite{sharp2019vector} showed that the gradient of this energy can be approximated efficiently by solving a sparse linear system. While their approach is robust and general, using only first-order derivatives precludes second-order convergence. Our analytical second-order derivatives of geodesic distance, in contrast, enable Newton-type minimization, which leads to significantly improved convergence (see Sec.~\ref{sec:result_Karcher_means}).
%
\subsection{Elastic Geodesic Triangles}
\label{sec:elastic_triangles}
%
In the previous examples, we considered potentials defined on geodesic networks. We now extend our formulation to model elastic membranes made of geodesic triangles, \ie, finite element-like triangles with geodesic edges. 
\par
For linear triangle finite elements in Euclidean geometry---also known as constant strain triangles---the deformation gradient maps edge vectors from the undeformed configuration to corresponding deformed edges as 
\begin{equation}
    \bF \bar{\be}_{ij}= \be_{ij} \ ,
\end{equation}
where $\bar{\be}_{ij}=(\bar{\bx}_j-\bar{\bx}_i)$ and ${\be}_{ij}=({\bx}_j-{\bx}_i)$ are undeformed and deformed edges, respectively.
Squaring both sides, we obtain
\begin{equation}
    \bar{\be}_{ij}^{\tran} \bC \bar{\be}_{ij} = |\be_{ij}|^2 \ , \quad \bC = \bF^{\tran}\bF \ ,
\end{equation}
where $\bC$ is the Cauchy-Green tensor. A direct translation of this deformation measure to geodesic triangles would require geodesic edges, expressed in a common coordinate system. This representation, however, is not available since the vertices of a geodesic triangle generally fall into different triangles of the hosting surface.  Fortunately, we can construct the Cauchy-Green tensor using only edge lengths (see inset figure). To this end, we first note that
\begin{equation}
\label{eqn:C_tensor}
    \bar{\be}_{ij}^{\tran} \bC \bar{\be}_{ij} = g_{ij}^2 \ , 
\end{equation}
\setlength{\columnsep}{3pt}
\begin{wrapfigure}[9] {r}{0.2\textwidth}
\vspace{-5pt}
    \centering
    \includegraphics[width=\linewidth]{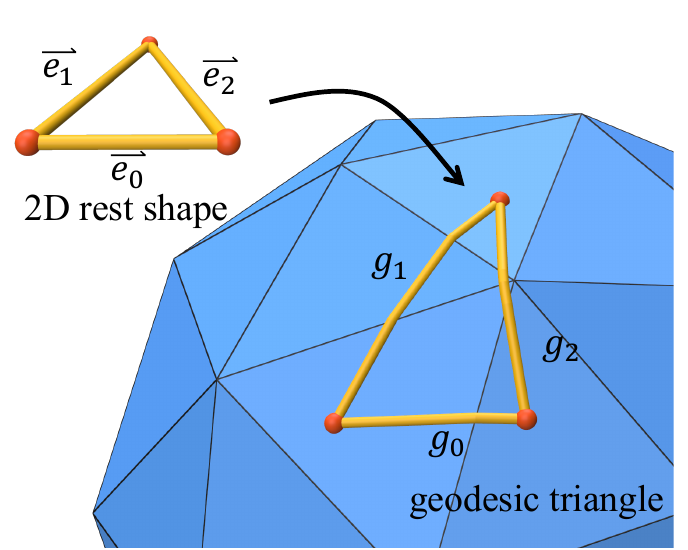}
    \label{fig:con_cur}
\end{wrapfigure}
for all three edges of a geodesic triangle. Using the fact that $\bC$ is a symmetric $2\times2$ tensor, it is evident that its three unknown entries can be computed by solving the three equations (\ref{eqn:C_tensor}). With this nonlinear deformation measurement at hand, we can then use a standard Neo-Hookean constitutive model for membrane elasticity. The corresponding energy density function is given as
\begin{equation}
    \Psi(\bC) = \frac{\mu}{2}(tr(\bC) - 2 - 2\log(J)) + \frac{\lambda}{2}\log(J)^2 \ ,
\end{equation}
where $J = \sqrt{\text{det}(\bC)}$. Since it only involves the lengths of geodesic edges, this energy density is a continuous function of the simulation variables $\bw$.

\paragraph{Discussion}
We assume that triangles are intrinsically flat when computing deformation gradients from edge lengths. While this is an approximation, it allows us to measure and penalize deformations in any direction, not just along edges. 
As another point, although it is possible to compute the area enclosed by a geodesic triangle, this
\setlength{\columnsep}{3pt}
\begin{wrapfigure}[6]{r}{0.25\textwidth} 
\vspace{-10pt}
    \centering
    \includegraphics[width=\linewidth]{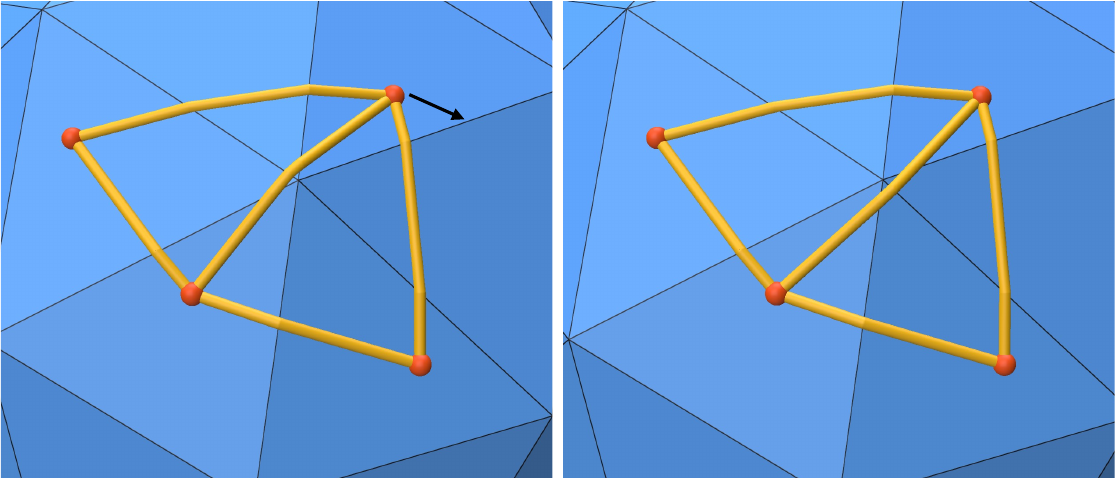}
    \label{fig:con_cur}
\end{wrapfigure}
area can be discontinuous as the geodesic path \textit{flips} in the vicinity of spherical vertices. An example is shown in the inset figure, where a small perturbation of one vertex changes the area of the two adjacent triangles abruptly. 
Instead of directly computing the area of geodesic triangles, our formulation captures changes in area through the determinant of the Cauchy-Green tensor, $\text{det}(\bC)$. Since computing $\bC$ requires only the lengths of geodesic edges, not their paths, we avoid discontinuities in area. 

%
\subsection{Two-way Coupling}
\label{sec:two_way}
The previous examples focused on geodesic elastic systems embedded in a rigid hosting surface. However, our approach can also be extended to full two-way coupling with deformable hosting objects. For these cases, the geodesic distance function takes the form
\begin{equation}
\label{eqn:geodesic_coupled}
    g(\bc(\bw, \bv), \bx(\bt(\bc(\bw, \bv), \bv)), \bv) \ ,
\end{equation}
where $\bv$ are the vertices of the hosting surface.
The first and second derivatives of this expression can again be obtained using sensitivity analysis; see App.~\ref{sec:two_way_coupling_derivatives}. 
We describe the coupled system through its potential energy,
\begin{equation}
    E_{\mathrm{coupling}}(\bq) = E_{\mathrm{network}}(\bq) + E_{\mathrm{host}}(\bq) \ ,
\end{equation}
where $\bq = (\bw,\bv)^{\tran}$ concatenates the barycentric coordinates of the embedded system and the vertices on the hosting surface. The first term $E_{\mathrm{network}}(\bq)$ models the energy of the embedded elastic system,
whereas $E_{\mathrm{host}}(\bq)$ is the elastic potential of the hosting object. We consider two types of models for the hosting object: a discrete shell model~\cite{grinspun2003discrete} and a standard volumetric finite element model~\cite{kim2020dynamic}.
\subsection{Differentiable Geodesic Voronoi Diagrams}
\label{sec:diff_gvd}
Up to this point, we have focused on elastic geodesic systems embedded in rigid and deformable hosting objects. We now extend our formulation to another application of intrinsic distances, geodesic Voronoi diagrams.
\paragraph{Geodesic Voronoi Diagrams}
For a given set of sample points (sites) on a surface, a surface Voronoi diagram partitions the surface mesh $\mathcal{M}$ into cells. A cell $C_i$ is the set of all points $\bx^*$ satisfying
\begin{equation}
    C_i = \{ \bx^* \in \mathcal{M} \ |\ g(\bx^*, \bs_i) < g(\bx^*, \bs_j), \ \forall j \neq i\} \ ,
\end{equation}
where $\bs_i$ and $\bs_j$ are the barycentric coordinates of two sites. The corresponding surface Voronoi diagram, using geodesic distance as the metric, is referred to as a geodesic Voronoi diagram (GVD).
%
To represent the Voronoi diagram, we use only the site locations $\bs$ as explicit degrees of freedom. The cell boundary vertices $\tbx$, which are either Voronoi vertices (with 3 or more adjacent cells) or intersections between Voronoi edges and mesh edges (with 2 adjacent cells), are given implicitly as the unique solution to the equidistance constraint
\begin{equation}
    g(\tbx_i, \bs_0) - g(\tbx_i, \bs_j) = 0, j=1\dots N_i-1
\end{equation}
where $N_i$ is the number of adjacent cells to cell boundary vertex $\tbx_i$ and $\bs_0,\dots,\bs_{N_i-1}$ are the sites of adjacent cells. We compute the geodesic Voronoi diagram by solving
\begin{equation}
\label{eqn:exact_condition}
    \tbx_i = \argminl_{\tbx^*}\: \sum_{j=1}^{N_i-1} (g(\tbx^*, \bs_0) - g(\tbx^*, \bs_j))^2 \
\end{equation}
for each boundary vertex. Note that this requires prior knowledge of cell connectivity, which we obtain using a fast GVD approximation~\cite{xin2022surfacevoronoi}; see also Algorithm~\ref{alg:exact_GVD}. 
%
%
%
\paragraph{Bi-level Optimization}
We now turn to optimizing geodesic Voronoi diagrams for higher-level objective functions. Consider a constrained optimization problem of the form
\begin{equation}
\label{eqn:bi_obj}
    \min_{\tbx, \bs} O(\tbx, \bs) \quad \text{s.t.} \quad \bff(\tbx,\ \bs) = \bZero \ ,
\end{equation}
where $O$ is an \textit{outer} design objective function and $\bff$ is the vector-valued equidistance constraint function given by the gradient of the \textit{inner} objective (\ref{eqn:exact_condition}).
%
%
The states $\tbx$ are coupled with the variables $\bs$ through the constraint and we write $\tbx = \tbx(\bs)$ to eliminate $\tbx$ as explicit degrees of freedom. 
Optimizing this objective with gradient-based methods requires the derivative $\frac{\dd \tbx}{\dd \bs}$, which we obtain through sensitivity analysis. We refer to App.~\ref{sec:dxds} for derivations. 
%
\par
The equidistance constraints derive from the definition of Vo\-ro\-noi diagrams, \ie, a point on a Voronoi edge must have the same distance to all adjacent sites. This includes Voronoi vertices as well as intersections between Voronoi edges and mesh edges, for which $\bff$ takes different forms.
For intersections between Voronoi edges and mesh edges,  $\tbx_i$ is equidistant to two sites and $\bff$ takes the form
\begin{equation}
    \bff_i(\tbx_i, \bs_0, \bs_1) = g(\tbx_i(\bs_0, \bs_1), \bs_0) - g(\tbx_i(\bs_0, \bs_1), \bs_1) \ .
\end{equation}
\begin{wrapfigure}{r}{0.15\textwidth} 
\vspace{-12pt}
    \centering
    \includegraphics[width=\linewidth]{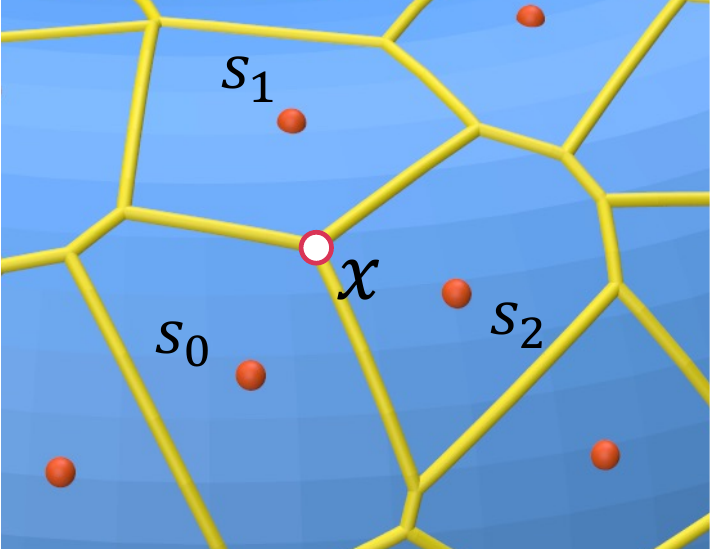}
\end{wrapfigure}
In this case, only one constraint equation is necessary to define a unique point along the intersected mesh edge, parameterized by a scalar as in Eq. (\ref{eqn:intersection_pt_parameterization}).
For Voronoi vertices that are equidistant to three or more sites (see inset), we use two constraint equations
\begin{equation}
\begin{aligned}
    \bff_{i,0}(\tbx_i, \bs_0, \bs_1) &= g(\tbx_i(\bs), \bs_0) - g(\tbx_i(\bs), \bs_1) \ , \\
    \bff_{i,1}(\tbx_i, \bs_0, \bs_2) &= g(\tbx_i(\bs), \bs_0) - g(\tbx_i(\bs), \bs_2) \ .
\end{aligned}
\end{equation}
\par
We minimize the inner objective (\ref{eqn:exact_condition}) using Newton's method and the outer design objective (\ref{eqn:bi_obj}) using quasi-Newton methods. A standard back-tracking line search is applied to both optimization processes.
%
\subsection{Implementation Details}
Our code is implemented in C++ using Eigen~\cite{eigenweb} for primary data structure and Intel TBB for parallelization. Linear systems are solved using CHOLMOD~\cite{chen2008algorithm}. We use the MMP algorithm~\cite{mitchell1987discrete} implemented in the Geometry Central Library~\cite{geometrycentral} for computing exact geodesics. 
Finally, we credit Polyscope~\cite{polyscope} for producing figures. Our code is available through the repository \url{https://github.com/liyuesolo/DifferentiableGeodesics}.

%% file: 4_Results.tex
\section{Results} 
\subsection{Energy-minimizing Geodesic Networks}
\label{sec:result_curve_networks}
\begin{figure}[h]
    \centering
    \includegraphics[width=\linewidth]{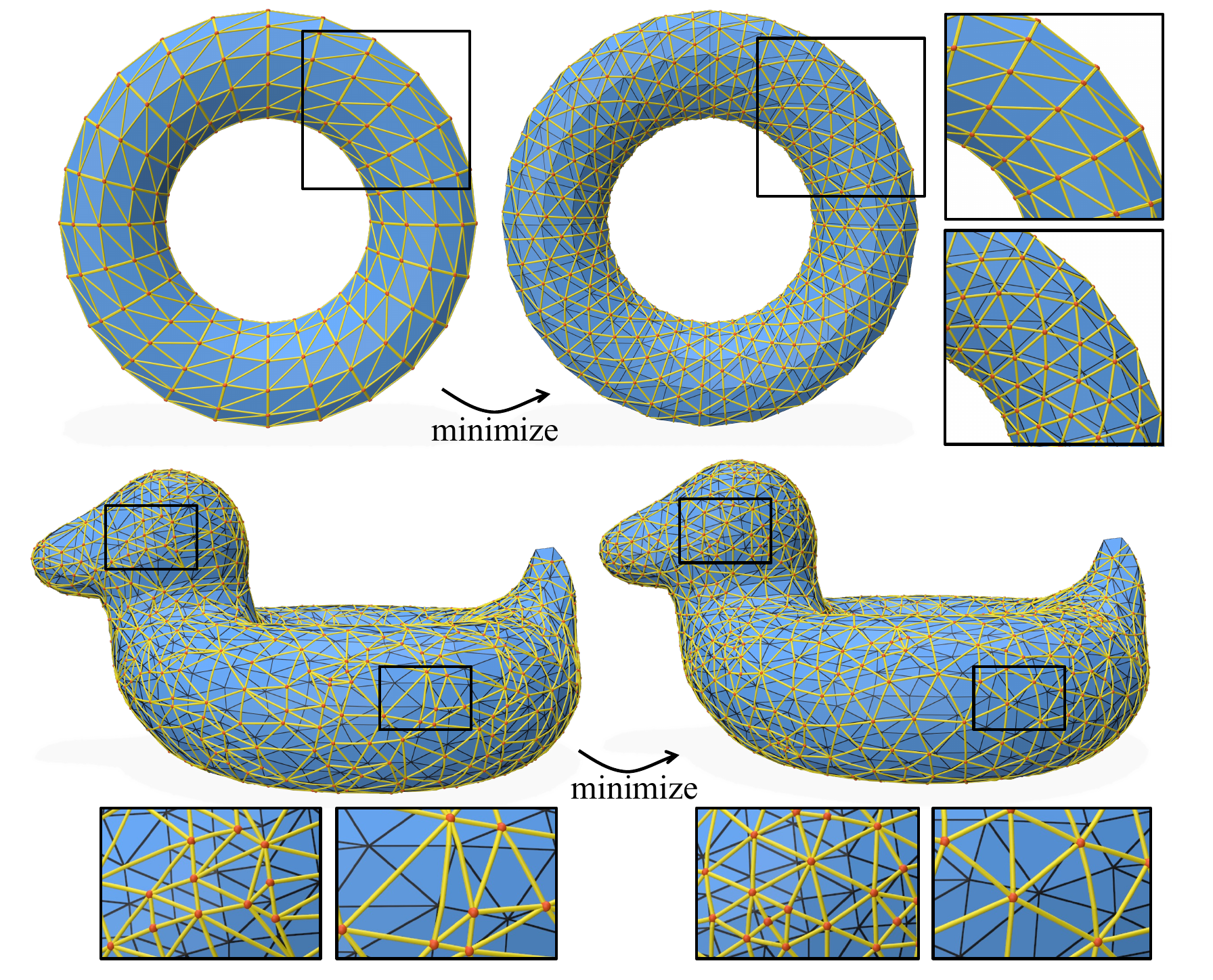}
    \caption{Elastic geodesic spring networks. We initialize the nodes (shown in red) in close proximity to either the vertices of the hosting mesh or its edge midpoints. Our method converges robustly in both scenarios.}
    \label{fig:geo_spring_torus}
\end{figure}

In the first set of examples (Fig.~\ref{fig:geo_spring_torus}), we simulate elastic curve networks comprised of zero-length geodesic springs. We showcase our approach on two challenging setups where the endpoints of the geodesic springs are positioned close to mesh vertices or edges. Our approach converges robustly in both scenarios. As can be seen in the close-up views, we arrive at energy minima that differ significantly from the initial network.
\begin{figure*}[ht!]
    \centering
    \includegraphics[width=\linewidth]{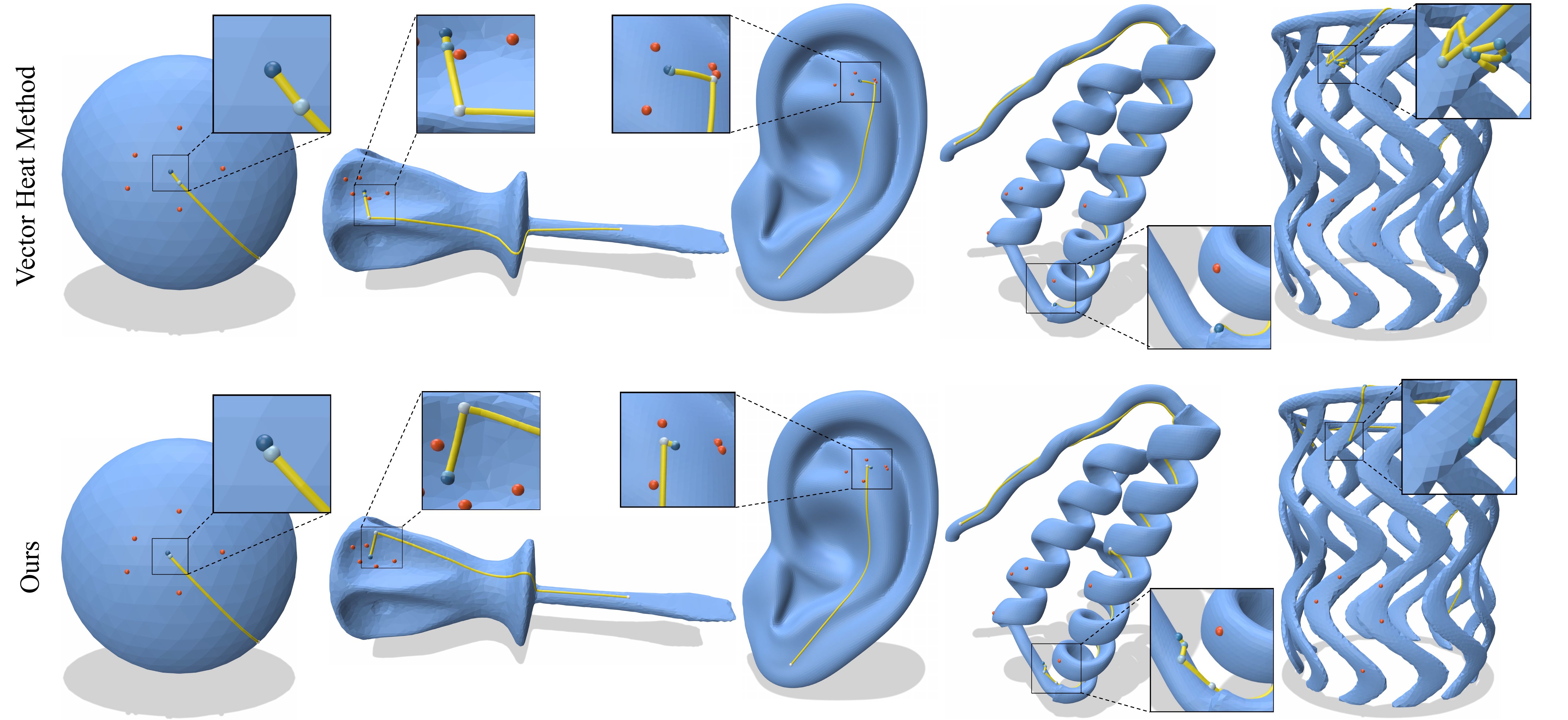}
    \caption{Qualitative comparisons with the Vector Heat Method~\cite{sharp2019vector} for computing Karcher Means on a selection of meshes. The yellow curves show the optimization trajectories toward the Karcher mean of a given set of points (shown in red). The initial guesses (chosen randomly) are shown in white and the Karcher mean in blue. The color gradient indicates steps toward the solution. As can be seen from these examples, our second-order solver allows for larger steps toward the minima (\textit{col 1-3}) and significantly fewer iterations to convergence (\textit{col 4-5}).}
    \label{fig:comparison_heat}
\end{figure*}
\begin{table*}
    \caption{Comparison to the Vector Heat Method~\cite{sharp2019vector} for computing Karcher Means. We report the runtimes and optimization statistics for the examples shown in Fig.~\ref{fig:comparison_heat}. \textit{Heat*} indicates using the convergence criterion in the official implementation~\cite{geometrycentral}, whereas \textit{Heat} denotes using the same criterion as ours. While the Vector Heat Method converges to visually acceptable solutions in similar numbers of iterations, their gradient-based update is limited to linear convergence. Leveraging a Newton solver with analytical second derivatives, our approach converges quadratically.}
    \centering
    \begin{tabular}
    {p{1.4cm}p{1.0cm}p{2.8cm}p{2.2cm}p{5.0cm}p{3.4cm}}
    \toprule
       Examples &  Triangles & Time[s] (Heat*/Heat/\textbf{Ours}) & Iterations (Heat*/Heat/\textbf{Ours}) &GradientNorm \quad\quad\quad\quad (Heat*/Heat/\textbf{Ours}) & Objective (Heat*/Heat/\textbf{Ours}) \\ 
       \midrule
        sphere & 1,280 & 0.0162 / 0.916 /  \textbf{0.0284} & 2 / 200 / \textbf{2} & 4.35 $\times 10^{-4}$/ 3.04 $\times 10^{-5}$ / \textbf{2.31} \boldmath$\times 10^{-11}$ & 0.0206 / 0.0204 / \textbf{0.00744} \\
        screwdriver & 6,786 & 0.119 / 17.5 /  \textbf{0.291} & 3 / 200 / \textbf{3} & 3.04 $\times 10^{-4}$ / 2.67 $\times 10^{-4}$ / \textbf{1.57} \boldmath$\times 10^{-11}$ & 0.00387 / 0.00383 / \textbf{0.00121}\\
        ear & 45,312 & 1.15 / 120 / \textbf{4.99}& 3 / 200 / \textbf{3} &1.13 $\times 10^{-5}$ / 1.85 $\times 10^{-6}$ / \textbf{3.20} \boldmath$\times 10^{-12}$ & 0.00974 / 0.00972 / \textbf{0.00225}\\
        protein & 60,820 & 17.0 / 120 / \textbf{12.8}& 25 / 200 / \textbf{8} & 1.81 $\times 10^{-1}$ / 9.64 $\times 10^{-2}$ / \textbf{3.05} \boldmath$\times 10^{-7}$ & 0.970 / 0.991 / \textbf{0.208}\\
        spiral cup & 34,874 & 0.576 / 19.3 /  \textbf{0.291}& 13 / 200 / \textbf{10} & 3.04 $\times 10^{-4}$ / 1.05 $\times 10^{-5}$ / \textbf{1.99} \boldmath$\times 10^{-11}$ & 1.34 / 1.58 / \textbf{0.280}\\
        \bottomrule
    \end{tabular}
    \label{tab:comparsion_vector_heat}
\end{table*}
\subsection{Karcher Means}
\label{sec:result_Karcher_means}
Our approach enables second-order optimization for computing Karcher means on triangle meshes. We compare our approach to the state-of-the-art 
Vector Heat Method~\cite{sharp2019vector} on a set of examples with different geometries and resolutions. While Sharp~\etal~\shortcite{sharp2019vector} propose an efficient method to compute the energy gradient by solving sparse linear systems, their approach is limited to linear convergence. As can be seen from the qualitative comparisons in Fig.~\ref{fig:comparison_heat}, our second-order solver enables larger steps and converges to the solution in significantly fewer iterations. 
We use the open-source reference implementation from the Geometry Central Library~\cite{geometrycentral}, which adopts a relative convergence criterion based on the current triangle area and step size. For our experiments, we impose an absolute convergence criterion based on the gradient norm. Details on the convergence criteria are given in App.~\ref{sec:convengence_criteria}. We report statistics for both convergence criteria in Table~\ref{tab:comparsion_vector_heat}, showing that our approach converges to much tighter tolerances with comparable performance. We set a maximum iteration of 200 for all approaches.
\par
Another baseline for computing Karcher means is provided by Mancinelli and Puppo~\shortcite{mancinelli2023computing}, which leverages approximate first and second derivatives of geodesic distance. While these approximations enable fast computations, the underlying convexity assumption for geodesic distance does not generally hold for arbitrary triangle meshes. 
Unfortunately, the publicly available algorithm\footnote{https://github.com/Claudiomancinelli90/RCM\_on\_meshes} did not yield successful outcomes for any of the problem instances shown in Fig.~\ref{fig:comparison_heat}.
We therefore conduct a simpler test, computing the Karcher mean of five randomly placed points on a spherical mesh (Fig.~\ref{fig:comparison_heat}, column 1). We repeat this experiment 100 times and report the average runtime, residual norm, and success rate in Table~\ref{tab:MancinellOUrs}. Despite being an order of magnitude slower, our approach ensures robust convergence across all test cases. 
\begin{table}[h!]
    \caption{Quantitative comparison with Mancinelli and Puppo~\shortcite{mancinelli2023computing} using 100 randomized problem instances on a spherical mesh. Averaged runtimes and gradient norms exclude failure cases.} 
    \centering
    \begin{tabular}{cccc}
    \toprule
        Method & Avg Time [s] & Avg |Grad| & Success Rate \\
        \midrule
        MancinelliPuppo & $9.54 \times 10^{-3}$ & $1.40 \times 10^{-2}$ & 17\%  \\
        \textbf{Ours} & \textbf{9.23} \boldmath$\times 10^{-2}$ & \textbf{6.52} \boldmath$\times 10^{-12}$ & \textbf{100\%} \\
        \bottomrule
    \end{tabular}
    \label{tab:MancinellOUrs}
\end{table}
\subsection{Energy-minimizing Geodesic Triangles}
\label{sec:result_elastic_triangle}
\begin{figure}[h]
    \centering
    \includegraphics[width=\linewidth]{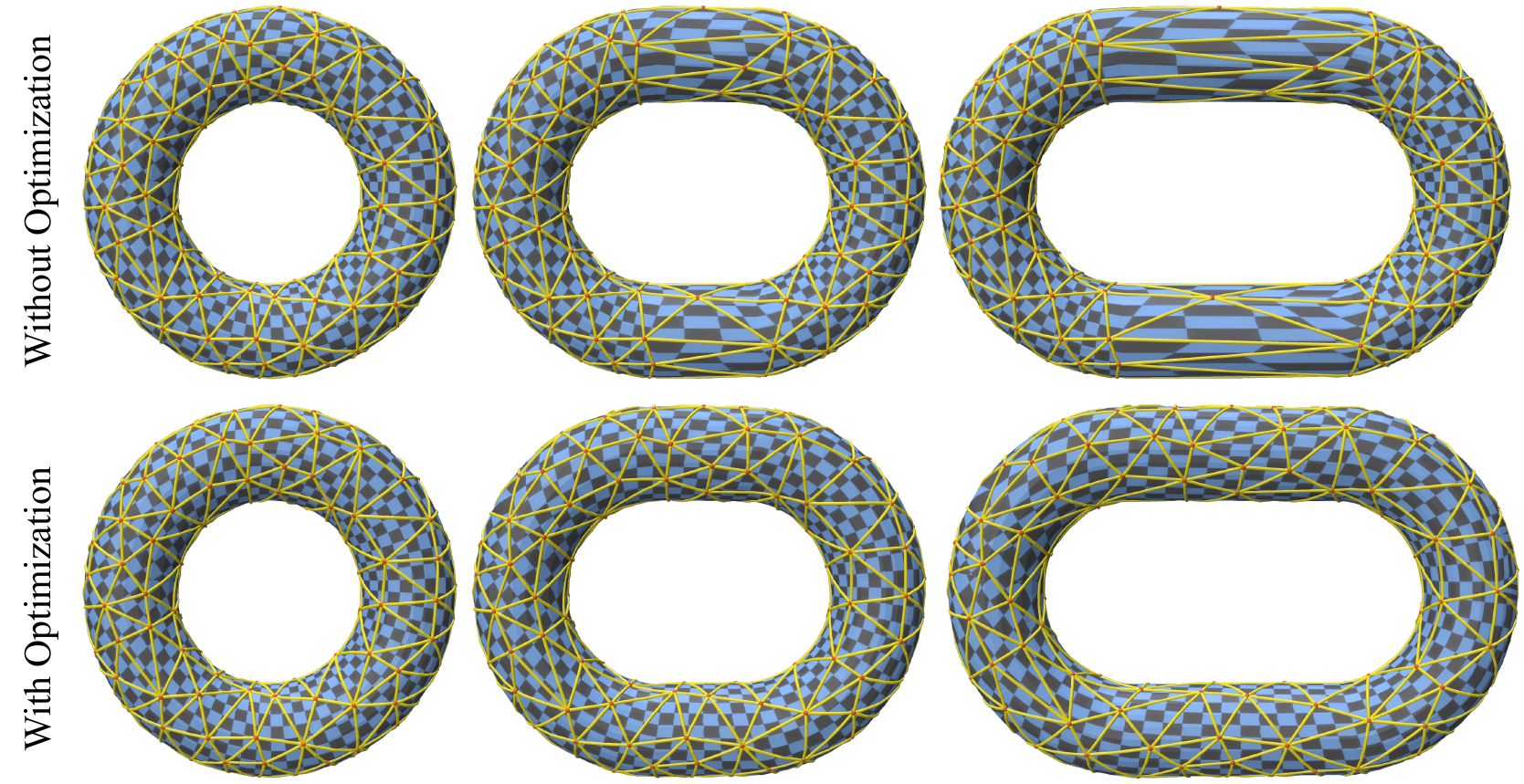}
    \caption{Simulation of an elastic membrane using geodesic triangles. We stretch the hosting torus mesh anisotropically to induce deformation of the embedded triangles. Keeping the (barycentric) membrane coordinates unchanged leads to large isolated distortions (\textit{top row}). Optimizing coordinates such as to minimize the membrane's energy leads to smoothly distributed deformations (\textit{bottom row}).
    }
    \label{fig:geo_tri_torus}
\end{figure}

We simulate membranes defined by elastic geodesic triangles. Inspired by previous work from Li \etal~\shortcite{li2013thin}, we simulate an elastic membrane made of geodesic triangles embedded in a rigid torus mesh (see Fig. \ref{fig:geo_tri_torus}).
The torus mesh is stretched along the horizontal axis of the image plane in a non-uniform way. Simply keeping the barycentric coordinates of the embedded membrane unchanged leads to large isolated distortions (Fig. \ref{fig:geo_tri_torus}, top row). Optimizing these coordinates such as to minimize the membrane's elastic energy yields smoothly distributed deformations (Fig. \ref{fig:geo_tri_torus}, bottom row). 
%
Whereas simulating this effect in a 2D parametric space requires UV unwrapping \cite{li2013thin}, our approach operates directly on the 3D surface mesh. Our Newton solver converges in fewer than 5 iterations on average for this sequence.
\subsection{Two-way Coupling}
\label{sec:result_two_way}
\begin{figure}[h]
    \centering
    \includegraphics[width=0.9\linewidth]{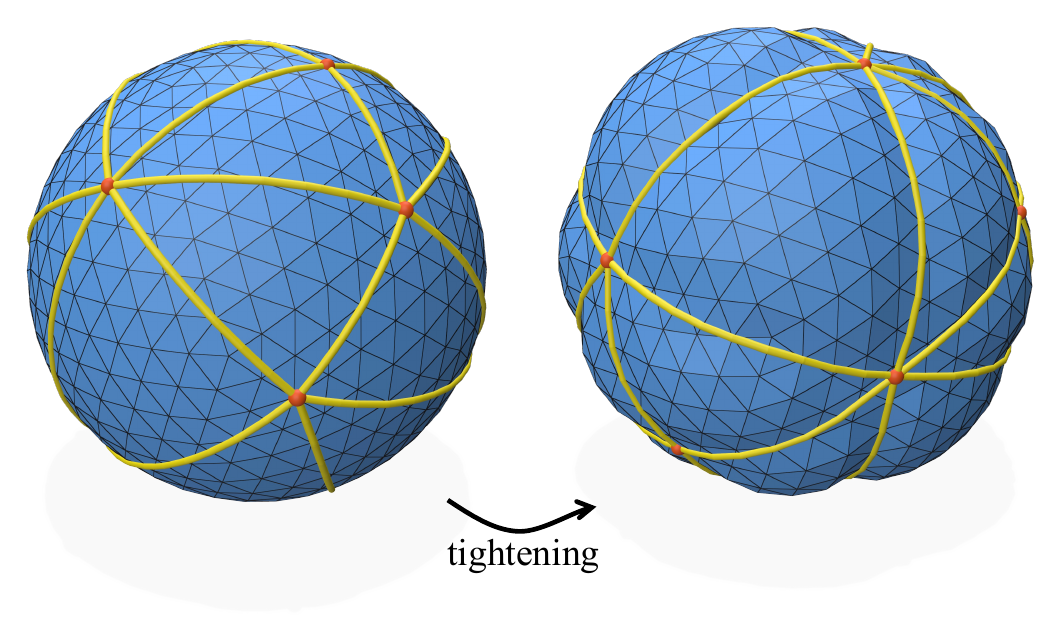}
    \caption{Two-way coupling. Simulation of a geodesic spring network on an inflated spherical shell. Realistic bulging effects emerge upon tightening the embedded curve network. }
    \label{fig:two_way_coupling}
\end{figure}

We now consider two-way coupling between a geodesic elastic network and its hosting surface. 
In the first example, we simulate the tightening of an elastic geodesic network embedded in an inflated spherical shell (see Fig. \ref{fig:two_way_coupling}). As the minimization proceeds, many regions of the shell bulge out in response to the compression forces induce by the embedded network. The hosting surface is simulated using a discrete shell model~\cite{grinspun2003discrete} augmented with a volume preservation term. To avoid factorizing a dense Hessian matrix resulting from the volume preservation term, we use the Sherman–Morrison formula~\shortcite{sherman1950adjustment} to compute the Newton step.
In Fig.~\ref{fig:teaser}, we show that our intrinsic minimization pipeline can be coupled with a volumetric mesh discretized using tetrahedron finite elements. The hosting elastic object is simulated using a Neo-Hookean constitutive model. The ears of the bunny bend down and the back bulges in response to the tightening of the geodesic spring network.
\subsection{Optimization of Geodesic Voronoi Diagrams}
\begin{figure}[h]
    \centering
    \includegraphics[width=\linewidth]{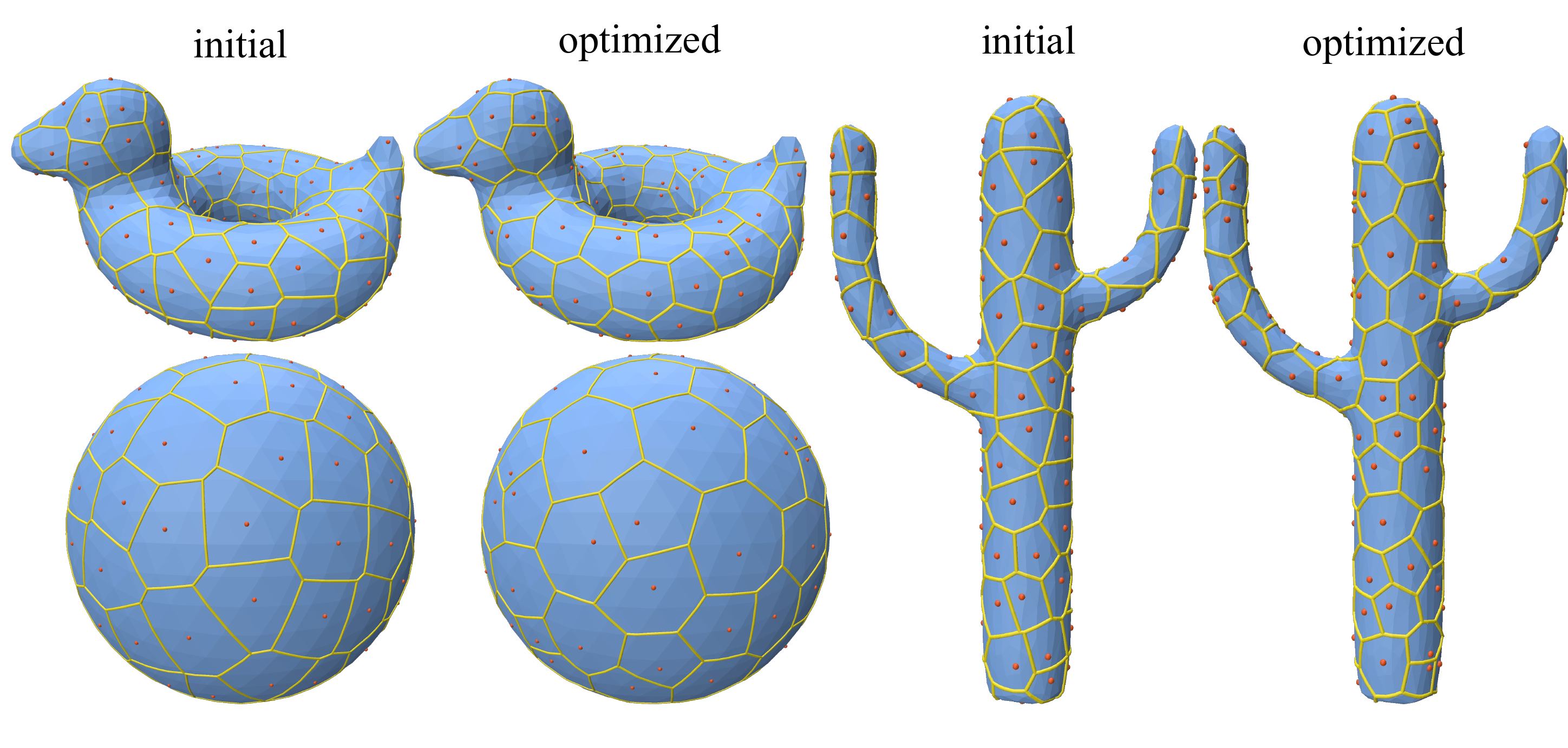}
    \caption{Optimizing for uniform edge length. By minimizing the deviation of all Voronoi edges from the target length, the optimized structure significantly reduces the length variation compared to the initial configuration.}
    \label{fig:similar_length}
\end{figure}

\begin{figure}[h]
    \centering
    \includegraphics[width=\linewidth]{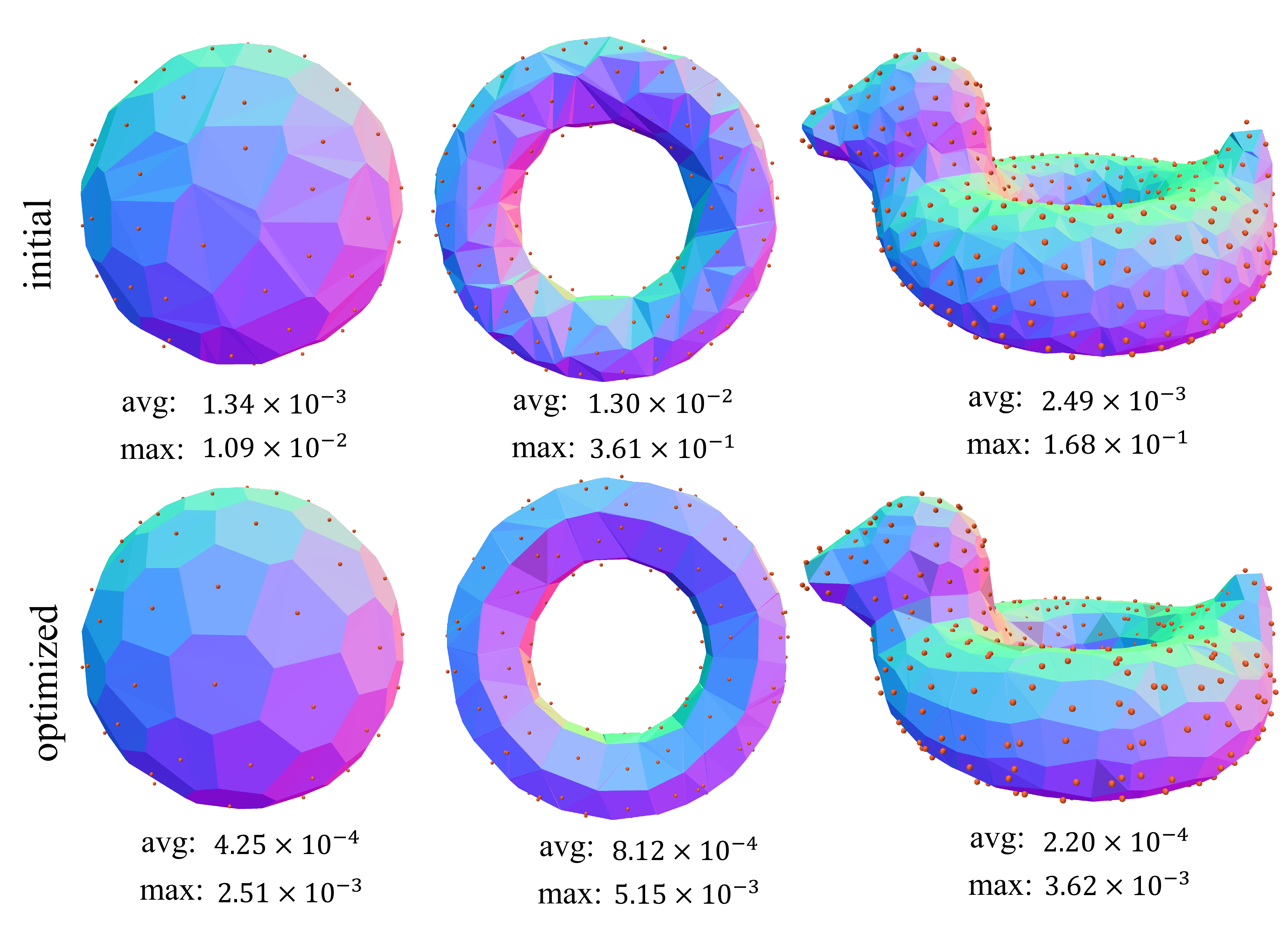}
    \caption{Optimizing for planarity. We optimize site locations such that the vertices of each Voronoi cell are as planar as possible. Our approach successfully reduces the average and maximum vertex-to-plane distance by an order of magnitude or more.}
    \label{fig:result_coplanrity}
\end{figure}

We consider three examples with different design objectives defined on geodesic Voronoi diagrams. Each example is formulated as a bi-level optimization problem using sensitivity analysis as described in section~\ref{sec:diff_gvd}.
\label{sec:result_diffGVD}
\paragraph{Length similarity}
In the first example, we optimize for uniform edge lengths such as to facilitate potential manufacturing with equal-length rods. To this end, we minimize the design objective  
\begin{equation}
\label{eqn:coplanar}
    \min_{\bs} O(\tbx(\bs), \bs) = \frac{1}{2} \sum_{(i,j)\in\mathcal{E}} (g(\tbx_i, \tbx_{j}) - L )^2 \ ,
\end{equation}
where  $\mathcal{E}$ is the index set of all Voronoi edges and $L$ is the target edge length, which we set to the average Voronoi edge length in the initial configuration. As can be seen in Fig.~\ref{fig:similar_length}, the optimized structures exhibit much less variation in edge length compared to the initial configurations.
\paragraph{Cell planarity} We consider the task of designing Voronoi diagrams with \textit{as-planar-as-possible} cells such that the 3D structure can be assembled from flat panels. We define a corresponding objective as
\begin{equation}
\label{eqn:coplanar}
    \min_{\bs} O(\tbx(\bs), \bs) = \frac{1}{2} \sum_{i=1}^{N} \sum_{j=1}^{N_i} \frac{1}{N_i} d(\tbx_j, \mathcal{P}_i(\tbx))^2 \ ,
\end{equation}
where $N$ denotes the total number of Voronoi vertices and $N_i$ is the number of vertices for Voronoi cell $i$. For each cell, we compute the least squares-fitting plane $\mathcal{P}_i$ and penalize the corresponding point-to-plane distance $d$ for each cell vertex.
As can be seen from Fig.~\ref{fig:result_coplanrity}, minimizing this planarity objective with our method leads to an order of magnitude reduction in average and maximum vertex-to-plane distance.

\begin{figure}[h]
    \centering
    \includegraphics[width=\linewidth]{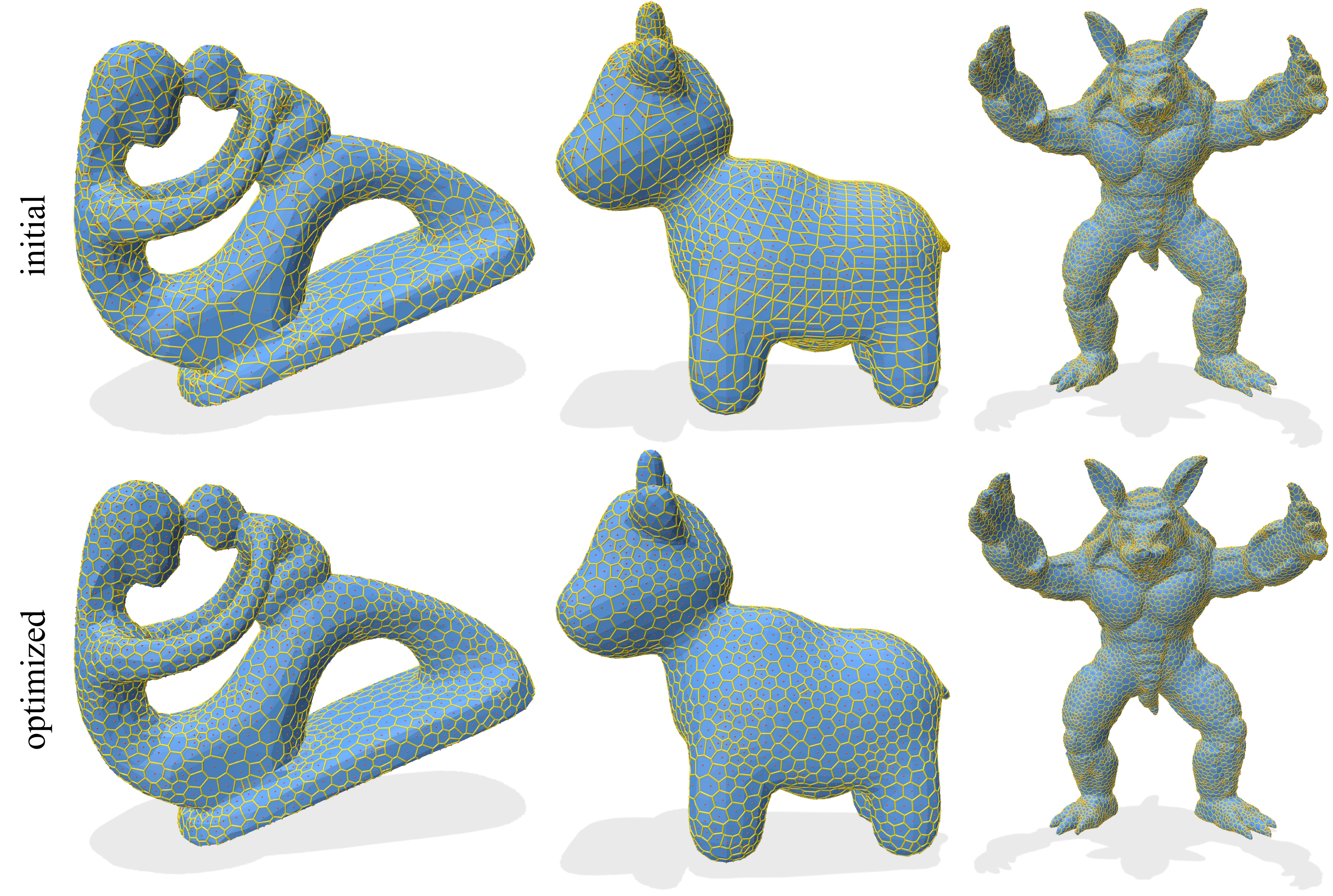}
    \caption{Optimizing cell regularity. Minimizing our regularity objective leads to quasi-isotropic cells when starting from a poorly shaped initial diagram.}
    \label{fig:GCVD_examples}
\end{figure}

\paragraph{Cell regularity}
Finally, we consider an objective to regularize cell shapes by minimizing the squared distances between each site and its surrounding vertices,
\begin{equation}
\label{eqn:gcvd}
    \min_{\bs} O(\tbx(\bs), \bs) = \frac{1}{2} \sum_{i=1}^{N} \sum_{j=1}^{N_i} g(\bs_i, \tbx_j(\bs))^2 \ .
\end{equation}
Fig.~\ref{fig:GCVD_examples} illustrates the effect of this objective on a set of Voronoi diagrams with irregular initial cell shapes. Despite the complexity of the hosting surfaces, the resulting cell shape distributions show significant improvements.
\subsection{Timings}
We report detailed simulation statistics and averaged timings for all examples. The timings listed in 
Tables~\ref{tab:comparsion_vector_heat}---\ref{tab:voro_timing} 
are measured on a workstation with an \textit{AMD Ryzen Threadripper PRO 5995WX} CPU. 
\begin{table*}[ht]
    \caption{Simulation statistics and average timings.}
    \centering
    \begin{tabular}{lcccrrrr}
    \toprule
      Example & Triangles & DoFs & Two-way & Exact Geodesics & Constructing Hessian & Solving Linear System & Newton Step\\
    \midrule
    Fig.~\ref{fig:teaser} & 9,396 & 20,838 & yes & 221.55 ms & 2.38 s & 690.8 ms & 3.64 s\\
    Fig.~\ref{fig:geo_spring_torus} torus & 576 & 1,152 & no & 73.43 ms & 48.47 ms & 42.58 ms & 172.31 ms\\
    Fig.~\ref{fig:geo_spring_torus} duck & 2,000 & 4,000 & no & 791.46 ms & 188.47 ms & 114.28 ms & 1.218 s\\
    Fig.~\ref{fig:comparison_heat} sphere & 1,280 & 2 & no & 8.19 ms & 112.5 $\mu$s& 63.5 $\mu$s& 12.52 ms\\
    Fig.~\ref{fig:comparison_heat} screwdriver & 6,786 & 2 & no & 63.20 ms & 12.24 ms & 43.67 $\mu$s & 94.47 ms\\
    Fig.~\ref{fig:comparison_heat} ear & 45,312 & 2 & no & 1.18 s & 45.86 ms & 59.6 $\mu$s & 1.65 s\\
    Fig.~\ref{fig:comparison_heat} protein & 60,820 & 2 & no & 710.72 ms & 948 ms & 40.86 $\mu$s & 1.668 s\\
    Fig.~\ref{fig:comparison_heat} spiral cup & 34,874 & 2 & no & 214.67 ms & 285.3 ms & 40.4 $\mu$s & 555.91 ms\\
    Fig.~\ref{fig:geo_tri_torus} & 2,304 & 282 & no & 89.36 ms & 5.91 ms & 596 $\mu$s & 96.75 ms\\
    Fig.~\ref{fig:two_way_coupling} & 1,280 & 30 & yes & 8.28 ms & 42.80 ms & 65.64 ms & 129.63 ms\\
    \bottomrule
    \end{tabular}
    \label{tab:timing_breakdown}
\end{table*}
\begin{table}[h]
    \caption{Statistics and average timing for energy-minimizing GVDs.}
    \centering
    \begin{tabular}{p{2.1cm}p{1.2cm}p{1.3cm}p{1.0cm}p{0.8cm}}
    \toprule
       Examples  &  Construct GVD[ms] & Time \quad /Iter [s]  & \#Tri & \#Sites \\
        \midrule
        Fig.~\ref{fig:similar_length} sphere & 5.87 & 1.90 & 1,280 & 98  \\
       Fig.~\ref{fig:similar_length} duck & 9.97 & 4.39  & 2,000 & 125 \\
        Fig.~\ref{fig:similar_length} cactus & 9.71 & 4.03 & 2,000 & 107 \\
        Fig.~\ref{fig:result_coplanrity} sphere & 14.37 & 0.474  & 1,280 & 396 \\
        Fig.~\ref{fig:result_coplanrity} torus & 6.41 & 0.0733 & 576 & 155  \\
        Fig.~\ref{fig:result_coplanrity} duck & 21.67 & 1.17 & 2,000 & 487 \\
        Fig.~\ref{fig:GCVD_examples} fertility & 66.46 & 7.09  & 2,250 & 2,250 \\
        Fig.~\ref{fig:GCVD_examples} cow & 93.12 & 39.3 & 5,856 & 2,928  \\
        Fig.~\ref{fig:GCVD_examples} armadillo & 262.6 & 180 & 20,000 & 10,000 \\
        \bottomrule
    \end{tabular}
    \label{tab:voro_timing}
\end{table}
\subsection{Ablation Study}
\label{sec:ablation_study}
\begin{figure}[h]
    \centering
    \includegraphics[width=\linewidth]{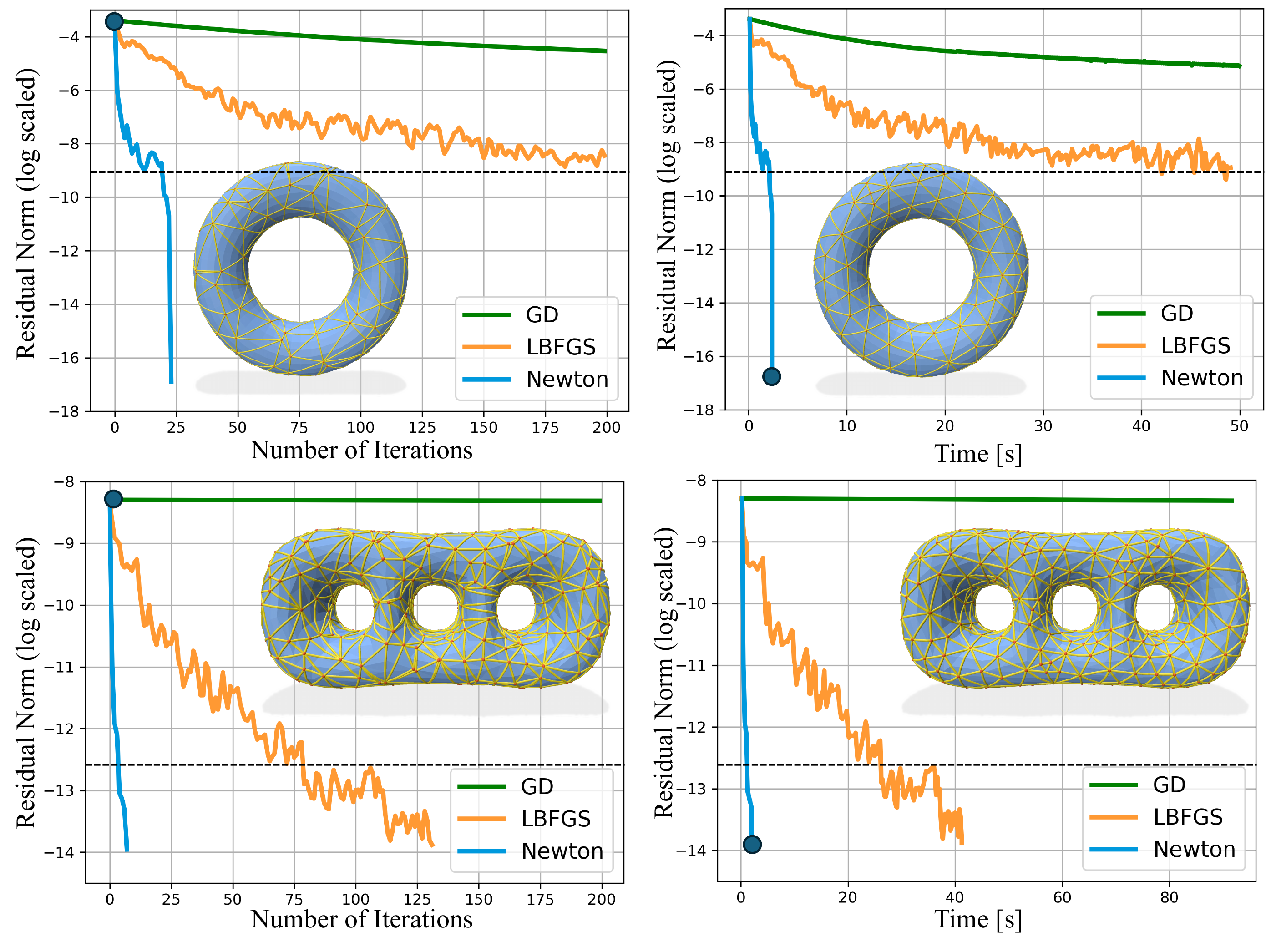}
    \caption{Convergence comparison. We use our second-order approach, gradient descent (GD), and L-BFGS to find equilibrium states of geodesic networks. Our approach exhibits quadratic convergence and demonstrates significantly better performance. Dashed lines indicate residual tolerances below which results become visually indistinguishable.
    Inset figures show the system states corresponding to the blue dots on the curves. }
    \label{fig:lbfgs}
\end{figure}

\begin{figure}[h]
    \centering
    \includegraphics[width=\linewidth]{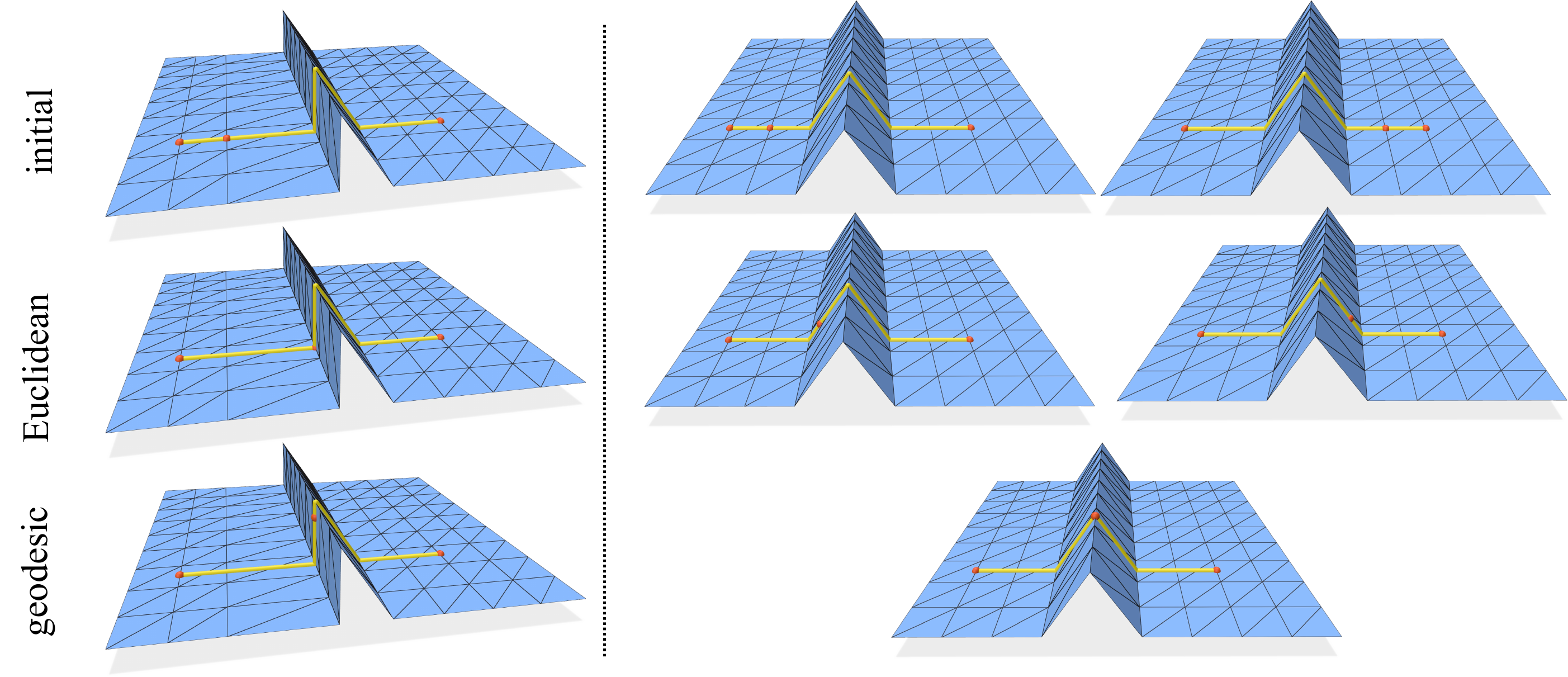}
    \caption{Spurious local minima when using Euclidean distance. In these two examples, we consider two zero-length springs connected by a free vertex and minimize the total elastic energy. Using Euclidean distance for the spring energy leads to undesired local minima while using geodesic distance resolves this problem. Geodesic paths are shown in all cases for visualization.}
    \label{fig:Euclidean1}
\end{figure}

\begin{figure}[h]
    \centering
    \includegraphics[width=\linewidth]{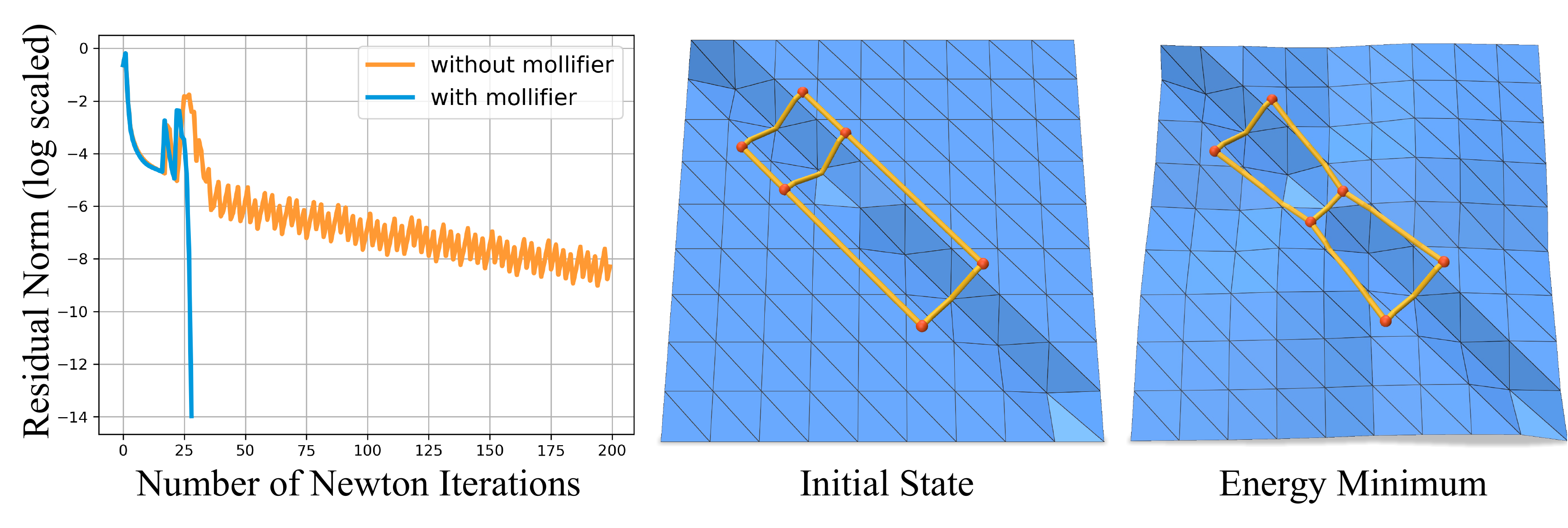}
    \caption{Mollification. In this two-way coupling example, the corners of the orange rectangle are fixed while the two inner vertices are free to slide on the surface. All embedded vertices are connected by geodesic springs with zero rest length. To arrive at the energy minimum, the inner spring must pass through the hyperbolic vertex of the hosting mesh. As can be seen from the convergence plot, without the mollifier (\textit{orange}), Newton's method does not achieve quadratic convergence due to the lack of $C^2$-continuity at the minimum. With the mollifier, however, our approach converges quadratically (\textit{blue}).}
    \label{fig:mollification}
\end{figure}

\paragraph{Comparison with L-BFGS}
We show that our analytical Hessian significantly improves convergence compared to first-order or quasi-Newton methods without sacrificing performance. We compare the convergence and performance of different methods on two energy minimization problems involving geodesic spring networks. As can be seen from Fig.~\ref{fig:lbfgs} (left), while L-BFGS offers better convergence than gradient descent (GD), Newton's method outperforms both by a large margin. In Fig.~\ref{fig:lbfgs} (right), we show that our second-order approach also offers significant performance gains. Thanks to the compact analytical expression of the second derivative, computing the analytical Hessian of the geodesic distance only takes a fraction of the overall computation time. The initial and optimized configurations can be found in the inset figures. Dashed lines in the figures indicate residual tolerances that lead to qualitatively on-par results.
We consider the results obtained from Newton and L-BFGS to be visually indistinguishable when the maximum difference in nodal position is smaller than $1\%$ of the scene bounding box diagonal. 
\paragraph{Euclidean Distance}
Using Euclidean distance instead of geodesic distance greatly simplifies computation but leads to additional and undesirable local minima in intrinsic simulation. We illustrate this problem on an embedded curve consisting of three vertices and two connecting segments modeled as zero-length springs (shown in Fig.~\ref{fig:Euclidean1}).
The endpoints are anchored to the hosting surface while the central vertex is free to move. We compare simulations using Euclidean and geodesic distance metrics. 
In the left example, the hosting surface exhibits a right angle, creating a local energy minimum for Euclidean-distance springs.
In the right example, multiple local minima exist for Euclidean springs and the eventual equilibrium configuration depends on the initial position of the central vertex. Using geodesic distance, however,  the simulation converges to the global optimum in both cases.
\paragraph{Mollification}
To study the influence of our smooth mollifier, we set up a two-way coupling example, where the geodesic path passes through a hyperbolic vertex in the minimum-energy state (Fig.~\ref{fig:mollification}, \textit{right}). In this example, a rectangle with an inner crossbar is embedded in a deformable shell. The four corner vertices of this rectangle are fixed to the surface and all edges are zero-length geodesic springs. In this case, energy minimization results in the inner edge passing through the hyperbolic vertex of the shell. As can be seen in the convergence plot to the left, without the mollifier, Newton's method struggles to converge even after 200 iterations. When the energy is mollified into a $C^2$-continuous function, Newton's method exhibits quadratic convergence.

%% file: 5_Conclusion.tex
\section{Conclusion \& Future Work}
We proposed a novel differentiable geodesic distance formulation for intrinsic minimization on triangle meshes. By applying the implicit function theorem to the variational formulation of shortest-path geodesics, we demonstrate that closed-form first and second derivatives can be obtained and significantly simplified, allowing for the use of Newton-type minimization solvers. We use our method to minimize a diverse set of objectives, including two-way coupling of embedded elastic structures with hosting geometries, Karcher means on arbitrary triangle meshes, and differentiable geodesic Voronoi diagrams.
\paragraph{Limitations}
While our intrinsic minimization paradigm enables robust second-order optimization on many instances of embedded elasticity, in almost all cases, the geodesics are under tension and the energy increases with geodesic length. However, when a geodesic curve is under compression and increasing length is energetically favorable, local distance maxima along the cut locus may lead to convergence failure. 
\par
We use the open-source implementation of the MMP algorithm for computing geodesic distance. While we perform this computation largely in parallel, it remains the computational bottleneck of our method. 
While our triangle elements based on geodesic edge lengths are promising for simulating embedded elastic membranes, the approximation accuracy decreases as the curvature within each geodesic triangle increases. Using denser geodesic triangulations can alleviate this issue to some extent, but using internal quadrature points and additional degrees of freedom could be a promising direction.
Finally, we only focused on triangle meshes in this work. Extending our intrinsic minimization paradigm to polygonal meshes is another avenue for future exploration.

%% file: Appendix.tex
\appendix
\section{First and Second Derivative of Geodesic Distance for Two-way Coupling}
\label{sec:two_way_coupling_derivatives}
The geodesic distance in this case is given by (\ref{eqn:geodesic_coupled}), which we also include here for completeness
\begin{equation}
    g(\bc(\bw, \bv), \bx(\bt(\bc(\bw, \bv), \bv)), \bv).
\end{equation}
The gradient of $g$ \wrt $\bw$ is
\begin{equation}
    \frac{\dd g}{\dd \bw} = \frac{\partial g}{\partial \bc}^{\tran} \frac{\partial \bc}{\partial \bw} + \frac{\partial g}{\partial \bx}^{\tran} \frac{\partial \bx}{\partial \bt} \frac{\partial \bt}{\partial \bc} \frac{\partial \bc}{\partial \bw},
\end{equation}
and gradient of $g$ \wrt $\bv$ is
\begin{equation}
    \frac{\dd g}{\dd \bv} = \frac{\partial g}{\partial \bc}^{\tran} \frac{\partial \bc}{\partial \bv} + \frac{\partial g}{\partial \bx}^{\tran} \frac{\partial \bx}{\partial \bt} (\frac{\partial \bt}{\partial \bc}\frac{\partial \bc}{\partial \bv} + \frac{\partial \bt}{\partial \bv}) + \frac{\partial g}{\partial \bx}^{\tran}\frac{\partial \bx}{\partial \bv}.
\end{equation}
Following the same procedure as detailed in Sec.~\ref{sec:dif_geo}, we again use the first-order optimality condition,
\begin{equation}
\label{eqn:implicit_geodesic_zero_coupled}
    \frac{\dd g}{\dd \bt} := \frac{\partial g}{\partial \bx}^{\tran} \frac{\partial \bx}{\partial \bt} = \bZero
\end{equation}
to simplify the gradient expression to
\begin{equation}
    \frac{\dd g}{\dd \bw} = \frac{\partial g}{\partial \bc}^{\tran} \frac{\partial \bc}{\partial \bw}, \quad \frac{\dd g}{\dd \bv} = \frac{\partial g}{\partial \bc}^{\tran} \frac{\partial \bc}{\partial \bv} + \frac{\partial g}{\partial \bx}^{\tran}\frac{\partial \bx}{\partial \bv}.
\end{equation}
The hessian is more complex due to the coupling. The complete expression include these individual blocks,
\begin{equation}
\begin{aligned}
    & \frac{\partial^2 g}{\partial \bw^2} = \frac{\partial \bc}{\partial \bw}^{\tran} \frac{\partial^2 g}{\partial \bc^2} \frac{\partial \bc}{\partial \bw} + \frac{\partial \bc}{\partial \bw}^{\tran} \frac{\partial^2 g}{\partial \bc \partial \bx} \frac{\partial \bx}{\partial \bt} \frac{\partial \bt}{\partial \bc}\frac{\partial \bc}{\partial \bw} + 
    \sum_i \frac{\partial g}{\partial c_i} \frac{\partial^2 c_i}{\partial \bw^2} \\
    & + \left(\frac{\partial \bx}{\partial \bt} \frac{\partial \bt}{\partial \bc}\frac{\partial \bc}{\partial \bw} \right)^{\tran} \left( \frac{\partial^2 g}{\partial \bc \partial \bx}\frac{\partial \bc}{\partial \bw} + \frac{\partial^2 g}{\partial \bx^2} \frac{\partial \bx}{\partial \bt} \frac{\partial \bt}{\partial \bc}\frac{\partial \bc}{\partial \bw} \right) \\
    & + 
    \left(\frac{\partial \bt}{\partial \bc} \frac{\partial \bc}{\partial \bw} \right)^{\tran}\sum_i \frac{\partial g}{\partial x_i} \frac{\partial^2 x_i}{\partial \bt^2} \left(\frac{\partial \bt}{\partial \bc} \frac{\partial \bc}{\partial \bw} \right)\\
    &+ 
    \frac{\partial \bc}{\partial \bw}^{\tran} \left( \sum_i \frac{\partial g}{\partial \bx} \frac{\partial \bx}{\partial t_i} \frac{\partial^2 t_i}{\partial \bc^2} \right)\frac{\partial \bc}{\partial \bw} + 
    \sum_i \frac{\partial g}{\partial \bx} \frac{\partial \bx}{\partial \bt}\frac{\partial \bt}{\partial c_i} \frac{\partial^2 c_i}{\partial \bw^2}.
\end{aligned}
\end{equation}
\begin{equation}
\begin{aligned}
    & \frac{\partial^2 g}{\partial \bv^2} = \frac{\partial \bc}{\partial \bv}^{\tran} \frac{\partial^2 g}{\partial \bc^2} \frac{\partial \bc}{\partial \bv} + \frac{\partial \bc}{\partial \bv}^{\tran} \frac{\partial^2 g}{\partial \bc \partial \bx} \left(\frac{\partial \bx}{\partial \bt} \left(\frac{\partial \bt}{\partial \bc}\frac{\partial \bc}{\partial \bv} + \frac{\partial \bt}{\partial \bv}\right)  + \frac{\partial \bx}{\partial \bv} \right)\\
    & + \sum_i \frac{\partial g}{\partial c_i}\frac{\partial^2 c_i}{\partial \bv^2}  + \left(\frac{\partial \bx}{\partial \bt} \left(\frac{\partial \bt}{\partial \bc}\frac{\partial \bc}{\partial \bv} + \frac{\partial \bt}{\partial \bv}\right)  + \frac{\partial \bx}{\partial \bv} \right)^{\tran} \frac{\partial^2 g}{\partial \bc \partial \bx}^{\tran} \frac{\partial \bc}{\partial \bv} \\ 
    &+ \left(\frac{\partial \bx}{\partial \bt} \left(\frac{\partial \bt}{\partial \bc}\frac{\partial \bc}{\partial \bv} + \frac{\partial \bt}{\partial \bv}\right) + \frac{\partial \bx}{\partial \bv} \right)^{\tran} \frac{\partial^2 g}{\partial \bx^2} \left(\frac{\partial \bx}{\partial \bt} \left(\frac{\partial \bt}{\partial \bc}\frac{\partial \bc}{\partial \bv} + \frac{\partial \bt}{\partial \bv}\right) + \frac{\partial \bx}{\partial \bv} \right)\\
    & + \sum_i (\frac{\partial \bt}{\partial \bc} \frac{\partial \bc}{\partial \bv} + \frac{\partial \bt}{\partial \bv})^{\tran} \frac{\partial g}{\partial x_i} \left(\frac{\partial^2 x_i}{\partial \bt^2} \left(\frac{\partial \bt}{\partial \bc}\frac{\partial \bc}{\partial \bv} +\frac{\partial \bt}{\partial \bv} \right) + \frac{\partial^2 x_i}{\partial \bt \partial \bv} \right)\\
    & + \frac{\partial g}{\partial \bx}^{\tran} \sum_i \frac{\partial \bx}{\partial t_i} \left( \frac{\partial \bc}{\partial \bv}^{\tran} \left( \frac{\partial^2 t_i}{\partial \bc^2}\frac{\partial \bc}{\partial \bv} + \frac{\partial^2 t_i}{\partial \bc \partial \bv} \right) + \sum_j \frac{\partial t_i}{\partial c_j} \frac{\partial^2 c_j}{\partial \bv^2} + \frac{\partial^2 t_i}{\partial \bc \partial \bv}\frac{\partial \bc}{\partial \bv} \right) \\
    & + \frac{\partial g}{\partial \bx}^{\tran} \sum_i \frac{\partial \bx}{\partial t_i} \left( \frac{\partial^2 t_i}{\partial \bv^2}\right) + \sum_i \frac{\partial g}{\partial x_i} \left( \frac{\partial^2 x_i}{\partial \bt \partial \bv} \left( \frac{\partial \bt}{\partial \bc}\frac{\partial \bc}{\partial \bv} + \frac{\partial \bt}{\partial \bv}\right)  + \frac{\partial^2 x_i}{\partial \bv^2}\right).
\end{aligned}
\end{equation}
\begin{equation}
\begin{aligned}
    &\frac{\partial^2 g}{\partial \bv \partial \bw} = \frac{\partial \bc}{\partial \bw}^{\tran} \left( \frac{\partial^2 g}{\partial \bc^2} \frac{\partial \bc}{\partial \bv} + \frac{\partial^2 g}{\partial \bc \partial \bx} \left(\frac{\partial \bx}{\partial \bt} \left(\frac{\partial \bt}{\partial \bc}\frac{\partial \bc}{\partial \bv} + \frac{\partial \bt}{\partial \bv}\right)  + \frac{\partial \bx}{\partial \bv} \right)  \right)\\
    &+  \left(\frac{\partial \bx}{\partial \bt}\frac{\partial \bt}{\partial \bc}\frac{\partial \bc}{\partial \bw}\right)^{\tran} \left(\frac{\partial^2 g}{\partial \bc \partial \bx} \frac{\partial \bc}{\partial \bv} + \frac{\partial^2 g}{\partial \bx^2} \left( \frac{\partial \bx}{\partial \bt} \left(\frac{\partial \bt}{\partial \bc}\frac{\partial \bc}{\partial \bv} + \frac{\partial \bt}{\partial \bv}\right) + \frac{\partial \bx}{\partial \bv} \right) \right)\\
    &+ \sum_i (\frac{\partial \bt}{\partial \bc} \frac{\partial \bc}{\partial \bw})^{\tran} \frac{\partial g}{\partial x_i} \left(\frac{\partial^2 x_i}{\partial \bt^2} \left(\frac{\partial \bt}{\partial \bc}\frac{\partial \bc}{\partial \bv} +\frac{\partial \bt}{\partial \bv} \right) + \frac{\partial^2 x_i}{\partial \bt \partial \bv} \right) \\
    &+ \frac{\partial \bc}{\partial \bw}^{\tran} \left(\frac{\partial g}{\partial \bx} \sum_i \frac{\partial \bx}{\partial t_i} \left( \frac{\partial^2 t_i}{\partial \bc^2}\frac{\partial \bc}{\partial \bv} + \frac{\partial^2 t_i}{\partial \bc \partial \bv} \right) \right)
    + \sum_i \frac{\partial g}{\partial \bx}^{\tran} \frac{\partial \bx}{\partial \bt} \frac{\partial \bt}{\partial c_i} \frac{\partial^2 c_i}{\partial \bv \partial \bw} \\
    &+ \sum_i \frac{\partial g}{\partial c_i} \frac{\partial c_i^2}{\partial \bv \partial \bw}.
\end{aligned}
\end{equation}
We begin by simplifying the hessian expression using (\ref{eqn:implicit_geodesic_zero_coupled}), and after removing the second derivatives of linear functions we obtain:
\begin{equation}
\label{eqn:d2gdw2_}
\begin{aligned}
    \frac{\partial^2 g}{\partial \bw^2} & = \frac{\partial \bc}{\partial \bw}^{\tran} \frac{\partial^2 g}{\partial \bc^2} \frac{\partial \bc}{\partial \bw} + \frac{\partial \bc}{\partial \bw}^{\tran} \frac{\partial^2 g}{\partial \bc \partial \bx} \frac{\partial \bx}{\partial \bt} \frac{\partial \bt}{\partial \bc}\frac{\partial \bc}{\partial \bw} \\
    & + \left(\frac{\partial \bx}{\partial \bt} \frac{\partial \bt}{\partial \bc}\frac{\partial \bc}{\partial \bw} \right)^{\tran} \left( \frac{\partial^2 g}{\partial \bc \partial \bx}\frac{\partial \bc}{\partial \bw} + \frac{\partial^2 g}{\partial \bx^2} \frac{\partial \bx}{\partial \bt} \frac{\partial \bt}{\partial \bc}\frac{\partial \bc}{\partial \bw} \right). \\
\end{aligned}
\end{equation}

\begin{equation}
\begin{aligned}
    & \frac{\partial^2 g}{\partial \bv^2} = \frac{\partial \bc}{\partial \bv}^{\tran} \frac{\partial^2 g}{\partial \bc^2} \frac{\partial \bc}{\partial \bv} + \frac{\partial \bc}{\partial \bv}^{\tran} \frac{\partial^2 g}{\partial \bc \partial \bx} \left(\frac{\partial \bx}{\partial \bt} \left(\frac{\partial \bt}{\partial \bc}\frac{\partial \bc}{\partial \bv} + \frac{\partial \bt}{\partial \bv}\right)  + \frac{\partial \bx}{\partial \bv} \right)\\
    & + \left(\frac{\partial \bx}{\partial \bt} \left(\frac{\partial \bt}{\partial \bc}\frac{\partial \bc}{\partial \bv} + \frac{\partial \bt}{\partial \bv}\right)  + \frac{\partial \bx}{\partial \bv} \right)^{\tran} \frac{\partial^2 g}{\partial \bc \partial \bx}^{\tran} \frac{\partial \bc}{\partial \bv} \\ 
    &+ \left(\frac{\partial \bx}{\partial \bt} \left(\frac{\partial \bt}{\partial \bc}\frac{\partial \bc}{\partial \bv} + \frac{\partial \bt}{\partial \bv}\right) + \frac{\partial \bx}{\partial \bv} \right)^{\tran} \frac{\partial^2 g}{\partial \bx^2} \left(\frac{\partial \bx}{\partial \bt} \left(\frac{\partial \bt}{\partial \bc}\frac{\partial \bc}{\partial \bv} + \frac{\partial \bt}{\partial \bv}\right) + \frac{\partial \bx}{\partial \bv} \right)\\
    &+ \sum_i \frac{\partial g}{\partial x_i} \left( \frac{\partial^2 x_i}{\partial \bt \partial \bv} \left( \frac{\partial \bt}{\partial \bc}\frac{\partial \bc}{\partial \bv} + \frac{\partial \bt}{\partial \bv}\right) \right) + \sum_i \frac{\partial g}{\partial x_i} \left( \frac{\partial^2 x_i}{\partial \bt \partial \bv} \left( \frac{\partial \bt}{\partial \bc}\frac{\partial \bc}{\partial \bv} + \frac{\partial \bt}{\partial \bv}\right) \right)^{\tran}.
\end{aligned}
\end{equation}
The mixed second-order partial derivative of $g$ \wrt $\bw$ and $\bv$ reads
\begin{equation}
\begin{aligned}
    &\frac{\partial^2 g}{\partial \bv \partial \bw} = \frac{\partial \bc}{\partial \bw}^{\tran} \left( \frac{\partial^2 g}{\partial \bc^2} \frac{\partial \bc}{\partial \bv} + \frac{\partial^2 g}{\partial \bc \partial \bx} \left(\frac{\partial \bx}{\partial \bt} \left(\frac{\partial \bt}{\partial \bc}\frac{\partial \bc}{\partial \bv} + \frac{\partial \bt}{\partial \bv}\right)  + \frac{\partial \bx}{\partial \bv} \right)  \right)\\
    &+  \left(\frac{\partial \bx}{\partial \bt}\frac{\partial \bt}{\partial \bc}\frac{\partial \bc}{\partial \bw}\right)^{\tran} \left(\frac{\partial^2 g}{\partial \bc \partial \bx} \frac{\partial \bc}{\partial \bv} + \frac{\partial^2 g}{\partial \bx^2} \left( \frac{\partial \bx}{\partial \bt} \left(\frac{\partial \bt}{\partial \bc}\frac{\partial \bc}{\partial \bv} + \frac{\partial \bt}{\partial \bv}\right) + \frac{\partial \bx}{\partial \bv} \right) \right)\\
    &+ \sum_i (\frac{\partial \bt}{\partial \bc} \frac{\partial \bc}{\partial \bw})^{\tran} \frac{\partial g}{\partial x_i} \frac{\partial^2 x_i}{\partial \bt \partial \bv} + \sum_i \frac{\partial g}{\partial c_i} \frac{\partial c_i^2}{\partial \bv \partial \bw}.
\end{aligned}
\end{equation}
\par
Now that all terms can be evaluated in close form except for $\frac{\partial \bt}{\partial \bc}$ and $\frac{\partial \bt}{\partial \bv}$ which can be computed using sensitivity analysis. 
We begin by differentiate both sides of (\ref{eqn:implicit_geodesic_zero}) \wrt $\bc$, 
\begin{equation}
\label{eqn:dgdtdc2}
    \frac{\partial \bx}{\partial \bt}^{\tran} \left( \frac{\partial^2 g}{\partial \bc \partial \bx} + \frac{\partial^2 g}{\partial \bx^2} \frac{\partial \bx}{\partial \bt} \frac{\partial \bt}{\partial \bc} \right) = \bZero.
\end{equation}
To obtain $\frac{\partial \bt}{\partial \bc}$ we rearrange the terms and solve for 
\begin{equation}
    \bA \frac{\partial \bt}{\partial \bc} = \bB, 
\end{equation}
where 
\begin{equation}
    \begin{aligned}
        &\bA = \frac{\partial \bx}{\partial \bt}^{\tran}\frac{\partial^2 g}{\partial \bx^2}\frac{\partial \bx}{\partial \bt}, \\
        &\bB = -\frac{\partial \bx}{\partial \bt}^{\tran}\frac{\partial^2 g}{\partial \bc \partial \bx}.
    \end{aligned}
\end{equation}
\par
Differentiating both sides of Eqn.~\ref{eqn:implicit_geodesic_zero_coupled} \wrt $\bv$, we arrive at
\begin{equation}
\begin{aligned}
\label{eqn:dgdtdv}
    &\frac{\partial \bx}{\partial \bt}^{\tran} \left( \frac{\partial^2 g}{\partial \bc \partial \bx}\frac{\partial \bc}{\partial \bv} + \frac{\partial^2 g}{\partial \bx^2}\left(\frac{\partial \bx}{\partial \bt} \left(\frac{\partial \bt}{\partial \bc}\frac{\partial \bc}{\partial \bv} + \frac{\partial \bt}{\partial \bv}\right)  + \frac{\partial \bx}{\partial \bv} \right) \right) \\
    & + \sum_i \frac{\partial g}{\partial x_i} \frac{\partial^2 x_i}{\partial \bt \partial \bv} \left( \frac{\partial \bt}{\partial \bc}\frac{\partial \bc}{\partial \bv} + \frac{\partial \bt}{\partial \bv}\right)  = \bZero.
\end{aligned}
\end{equation}
To obtain $\frac{\partial \bt}{\partial \bv}$ we again rearrange the terms and solve for 
\begin{equation}
    \bC \frac{\partial \bt}{\partial \bv} = \bD, 
\end{equation}
where 
\begin{equation}
    \begin{aligned}
        &\bC = \frac{\partial \bx}{\partial \bt}^{\tran}\frac{\partial^2 g}{\partial \bx^2}\frac{\partial \bx}{\partial \bt}, \\
        &\bD = -\frac{\partial \bx}{\partial \bt}^{\tran}\frac{\partial^2 g}{\partial \bc \partial \bx}\frac{\partial \bc}{\partial \bv} -\frac{\partial \bx}{\partial \bt}^{\tran}\frac{\partial^2 g}{\partial \bx^2}\left(\frac{\partial \bx}{\partial \bt}\frac{\partial \bt}{\partial \bc}\frac{\partial \bc}{\partial \bv} + \frac{\partial \bx}{\partial \bv}\right) - \sum_i \frac{\partial g}{\partial x_i}\frac{\partial^2 x_i}{\partial \bt \partial \bv}.
    \end{aligned}
\end{equation}

Multiply (\ref{eqn:dgdtdc2}) by $(\frac{\partial \bt}{\partial \bc}\frac{\partial \bc}{\partial \bw})^{\tran}$ on the left and $\frac{\partial \bc}{\partial \bw}$ on the right we have
\begin{equation}
\label{eqn:simplify_dgdtdc_again}
\begin{aligned}
    \left( \frac{\partial \bx}{\partial \bt}\frac{\partial \bt}{\partial \bc}\frac{\partial \bc}{\partial \bw} \right)^{\tran} \left( \frac{\partial^2 g}{\partial \bc \partial \bx}\frac{\partial \bc}{\partial \bw} + \frac{\partial^2 g}{\partial \bx^2} \frac{\partial \bx}{\partial \bt}\frac{\partial \bt}{\partial \bc}\frac{\partial \bc}{\partial \bw} \right) &= \bZero.
\end{aligned}
\end{equation}
Therefore, as in the one-way coupling case, $\frac{\partial^2 g}{\partial \bw^2}$ can be further simplified to
\begin{equation}
\begin{aligned}
    \frac{\partial^2 g}{\partial \bw^2} & = \frac{\partial \bc}{\partial \bw}^{\tran} \frac{\partial^2 g}{\partial \bc^2} \frac{\partial \bc}{\partial \bw} + \frac{\partial \bc}{\partial \bw}^{\tran} \frac{\partial^2 g}{\partial \bc \partial \bx} \frac{\partial \bx}{\partial \bt} \frac{\partial \bt}{\partial \bc}\frac{\partial \bc}{\partial \bw}. \\
\end{aligned}
\end{equation}
\section{Convergence Criteria}
\label{sec:convengence_criteria}
Here we provide the details for computing the convergence criteria shown in Fig.~\ref{fig:comparison_heat}.
Let us use $\frac{\dd O}{\dd \bw}^{i}$ to define the gradient of the energy shown in (\ref{eqn:karcher}) \wrt the barycentric coordinate $\bw$ of the Karcher mean point in iteration $i$, and the convergence criterion defined in Geometry Central is given by
\begin{equation}
    \left|\alpha \cdot \frac{\dd O}{\dd \bw}\right| < \frac{1}{3} \sqrt{\mathcal{A}^i},
\end{equation}
where $\mathcal{A}$ is the area of the triangle that the Karcher mean point lies on in iteration $i$, and $\alpha$ is the line search scaling factor to ensure energy decrease.
\par
Our absolute convergence criterion is defined as
\begin{equation}
    \left|\frac{\dd O}{\dd \bw}\right| < 10^{-6}.
\end{equation}
As a reference of scale, for a given input geometry, we perform a uniform scaling such that the diagonal of its bounding box is unit length. While we set the tolerance to be $10^{-6}$, our approach often arrives at values orders-of-magnitude smaller thanks to the quadratic convergence.
\section{Geodesic Voronoi Diagram}
\label{sec:dxds}
In this section, we demonstrate how to leverage sensitivity analysis to compute the gradient of the objective function defined on the GVDs, \ie, (\ref{eqn:bi_obj}) in Sec.~\ref{sec:diff_gvd}. The gradient of this objective is given by
\begin{equation}
\label{eqn:dods_GVD}
    \frac{\dd O}{\dd \bs} = \frac{\partial O}{\partial \bs} + \frac{\partial O}{\partial \tbx}^{\tran} \frac{\dd \tbx}{\dd \bs}.
\end{equation}
Whereas $\frac{\partial O}{\partial \bs}$ and $\frac{\partial O}{\partial \tbx}$ can be computed algebraically, $\frac{\dd \tbx}{\dd \bs}$ requires the following sensitivity analysis
\begin{equation}
\begin{aligned}
    \frac{\dd \bff}{\dd \bs} = \bZero &= \frac{\partial \bg}{\partial \bs} + \frac{\partial \bff}{\partial \tbx}^{\tran} \frac{\dd \tbx}{\dd \bs},\\
    \frac{\dd \tbx}{\dd \bs} &= -\frac{\partial \bff}{\partial \tbx}^{-1}\frac{\partial \bff}{\partial \bs}.
\end{aligned}
\end{equation}
\paragraph{Pseudocode}
Alg.~\ref{alg:exact_GVD} provides pseudocode for computing the exact geodesic Voronoi diagram. Alg.~\ref{alg:min_gvd} gives the general recipe we follow for minimizing objectives defined on GVDs.
\begin{algorithm}[h]
\caption{Compute Exact Geodesic Voronoi Diagram}\label{alg:exact_GVD} 
\KwData{triangle mesh $\mathcal{M}$, number of Voronoi sites $n$, Voronoi site location $\bs$}
\KwResult{Voronoi boundary vertex position $\tbx$}
$\tbx^*$ $\gets$ ComputeGVDApproximation($\mathcal{M},n,\bs$) \quad // ~\cite{xin2022surfacevoronoi}\\
\textbf{parallel} \For{$i\gets 1$ \KwTo \text{len} $(\tbx^*)$}{
    $\tbx_i$ $\gets$ ComputeExactGVD($\mathcal{M},n,\bs, \tbx^*$) \quad // Eqn.(\ref{eqn:exact_condition})\\
}
\end{algorithm}
\begin{algorithm}
\caption{Energy-minimizing Geodesic Voronoi Diagram}\label{alg:min_gvd} 
\KwData{triangle mesh $\mathcal{M}$, number of Voronoi sites $n$, Voronoi site location $\bs$}
\KwResult{Voronoi site location $\bs^*$}
$\bs^* \gets \bs$, $\epsilon \gets 10^{-6}$ // initialize
\\
    \While{\textit{True}}
    {
        $\tbx$ $\gets$ ComputeExactGVD($\mathcal{M},n,\bs^*$) \quad\quad // Alg.(\ref{alg:exact_GVD})\\
        $\nabla O \gets$ computeGrad($\mathcal{M}, n, \bs^*, \tbx$) \quad\quad // Eqn.(\ref{eqn:dods_GVD})\\
        \If{$|\nabla O| < \epsilon$}{return $s^*$\;} 
        $O_0 \gets$ computeObj($\mathcal{M}, n, \bs^*, \tbx$) \quad\quad // Eqn.(\ref{eqn:gcvd})\\
        $\nabla^2 O \gets$ computeHessApprox($\mathcal{M}, n, \bs^*, \tbx$) \ \\
        $\triangle \bs \gets -(\nabla^2 O)^{-1} \nabla O$ \quad\quad\quad\quad\quad// linear solve  \\
        $\alpha \gets 1$  \quad\quad\quad\quad\quad\quad\quad\quad\quad // line search step size\\
        \While{\textit{True}}
        {
            $\bs^* \gets \bs^* + \alpha \cdot \triangle \bs$ \quad\quad\quad\quad\quad\quad\quad\quad\quad // Eqn.(\ref{eqn:tracing}) \\
            $\tbx$ $\gets$ ComputeExactGVD($\mathcal{M},n,\bs^*$) \quad\quad // Alg.(\ref{alg:exact_GVD})\\
            $O_1 \gets$ computeObj($\mathcal{M}, n, \bs^*, \tbx$) \quad\quad\quad // Eqn.(\ref{eqn:gcvd})\\
            \If{$O_1 < O_0$}{break\;} 
            $\alpha \gets 0.5 \cdot \alpha$
        }
    }
\end{algorithm}